\documentclass[times,12pt]{article}
\pdfoutput=1
\usepackage{amsmath,amssymb,amsfonts,latexsym,amsthm,enumerate,url}
\usepackage{graphicx}
\usepackage{mathrsfs}
\usepackage{xcolor}
\usepackage{float}
\usepackage{mathtools}
\usepackage{adjustbox}
\usepackage{graphicx}
\date{}

\newtheorem{thm}{Theorem}[section]

\newtheorem{prop}[thm]{Proposition}

\newcommand{\beq}[1]{\begin{equation}\label{#1}}
\newcommand{\enq}[0]{\end{equation}}

\newcommand{\remove}[1]{}

\newcommand{\comment}[1]{}

\title{Random Circuit Sampling: Fourier Expansion and Statistics}
\author {Gil Kalai, Yosef Rinott, and Tomer Shoham}

\begin{document}
        \maketitle

\begin{abstract}

Considerable effort in experimental quantum computing
    is devoted to noisy intermediate scale quantum
  computers (NISQ computers). Understanding the effect of noise is important
  for various aspects of this endeavor including notable claims for
  achieving quantum supremacy and attempts to demonstrate quantum error correcting codes. 
  In this paper we use Fourier methods combined with statistical analysis
  to study the effect of noise. In particular, we use Fourier analysis to refine the linear cross-entropy 
  fidelity estimator. We use both analytical methods and simulations to study the effect of readout and gate errors, and we use our analysis to study the samples of Google's 2019 quantum supremacy experiment. 

\end{abstract}



\section{Introduction}\label{sec:Int}
The 2019 paper ``Quantum supremacy using a programmable superconducting processor'' \cite{Aru+19} described an experiment carried out by Google's Sycamore quantum computer.  
Google's Sycamore quantum computer performed a \textit{sampling task}; that is,
it generated (with considerable noise) random bitstrings (vectors of zeroes and ones) of length $n$, $12 \le n\le 53$,
from certain discrete probability distributions supported on all such $2^n$ bitstrings. 
Google's  sampling task is referred to as random circuit sampling (RCS). The samples are from  probability distributions that depend on the circuits (a circuit can be regarded as a program that the quantum computer runs); These probability distributions are not computed by the quantum computers, and for large number of qubits their computation is claimed to be infeasible (even for a quantum computer). 
In the Google experiment the Sycamore quantum computer produced samples, each 
consisting 
of several hundred thousand binary strings (``bitstrings"), for each of around 1000 random circuits. For each sample, a certain statistic  called the {\it linear cross-entropy fidelity estimator} denoted by ${\mathcal F}_{XEB}$ (and sometimes simply by $XEB$) was computed. Based on the value of this estimator it was concluded that the samples produced by the quantum computer represent ``quantum supremacy."     

This paper continues our statistical analysis  
of the data and the statistical methods 
of the Google experiment and may have wider interest for the study of NISQ computers. For our earlier works, see 
\cite {RSK22,KRS22d,KRS23}. In this paper we rely, for the first time, on simulations for noisy quantum circuits, and for that purpose we used both the Google and the IBM simulators. (We also ran a few 5-qubit experiments on freely available IBM 7-qubit quantum computers.) All this experimental data and simulations are based on samples of $N=500,000$ bitstrings for every circuit that we study. In the appendix we also use our statistical tools to analyze samples for 12-qubit computation from a recent experiment \cite {Blu+23} of logical circuits based on neutral atoms.

The main novelty of this paper is the use of Fourier analysis (the Fourier--Walsh expansion) for statistical analysis of samples coming from NISQ computers and from simulations and, in particular, to refine the ${\cal F}_{XEB}$ fidelity estimator according to Fourier degrees. Put 
$[n]=\{1,2,\dots,n\}$. Recall that every real function $f$ defined on 0-1 vectors $x$ of length $n$, the Fourier--Walsh expansion of $f$ 
$$f(x)=\sum \{\widehat f(S)W_S(x):S \subset [n]\},$$
is a representation of $f$ as a linear combination of the Fourier--Walsh functions $W_S(x)$, where $S$ goes over all subsets of $[n]$. (See section \ref{s:fourier} for details.)
The coefficients $\widehat f(S)$ are referred to as the Fourier--Walsh coefficients of $f$, and if $k=|S|$ then $\widehat f(S)$ is referred to as a degree-$k$ Fourier--Walsh coefficient. 
We use both analytical methods and simulations to study the effect of readout and gate errors on the refined Fourier-based estimators, and use this study to examine the data from the Google experiment and from simulations. 

For an early work on noise stability, noise sensitivity and Fourier--Walsh expansion see  Benjamini, Kalai, and Schramm \cite {BKS99}.\footnote {The Fourier--Walsh expansion is also related to the ANOVA decomposition in statistics that goes back to Hoeffding, see, e.g.,  \cite {KarRin82}.} Kalai and Kindler \cite {KalKin14} used a related noise operator based on 
Fourier--Hermite expansion 
to study noisy boson sampling.
In the context of random circuit sampling, Fourier--Walsh expansion played a role in Boixo et al. \cite {BSN17}, Gao and Duan \cite {GaoDua18}, Kalai \cite {Kal18, Kal20}, 
and Aharonov et al. \cite {AGLLV22}.

A basic insight in the analysis of Boolean functions is the relevance of the noise operator $T_{\rho}$, where $\rho$ is a real number, $0 \le \rho\le 1$, defined as follows. 
$$T_{\rho}(f) = \sum \{\rho^{|S|} \widehat f(S)W_S(x):S \subset [n]\}.$$
Readout errors, namely, reading with probability $q$ a qubit as ``0" when it is in fact ``1" and vice versa, directly correspond to the noise operator $T_{(1-2q)}$. When the probability distribution described by a random circuit $C$ on $n$ qubits is given by the function ${\cal P}_C(x)$, one ingredient of our analysis is finding the best fit to the noisy data of the form
\begin {equation}
\label {e:noise-model-rho}
s T_{(1-2q)} ({\cal P}_C(x)) + \frac{1-s}{2^n},
\end {equation}
where $s,q$ are real numbers, $0 \le s \le 1$ and $0 \le q \le 1/2$. As it turned out (and can be anticipated from theoretical works \cite {Gao+21,AGLLV22}), the value of $q$ may reflect not only the simple connection with readout errors but also  a subtle effect of gate errors.  We will refer to the value of $q$ that best fits the data as the ``effective readout error." The effect of gate errors towards larger effective readout errors is witnessed in data coming from Google's simulator but not in the samples of the Google quantum supremacy experiment.


Here is a brief summary of this paper. Section \ref {s:back} provides background on Fourier--Walsh analysis and on the Google quantum supremacy experiment \cite{Aru+19}. Section \ref {s:rcs-four} describes the Fourier--Walsh distributions coming from random circuit sampling. Section \ref{s:stat-fourier} develops statistical estimates that refine the ${\cal F}_{XEB}$ estimate (and related estimates) to study the Fourier behavior. Section \ref {s:gates} applies these estimates to study the samples of the Google experiment as well as samples from simulations. Of special interest to us is the effect of gate errors on the Fourier--Walsh coefficients. Section \ref {s:fidelity} studies fidelity estimators. We extend some of our analysis from \cite {RSK22} to data coming from simulations of noisy quantum circuits and we concentrate on the estimators based on readout errors introduced in Section 6 of \cite {RSK22} which are related to the Fourier--Walsh behavior.  Section \ref {s:conc} concludes. 


\newpage

\section {Background} 
\label {s:back}

\subsection {Fourier--Walsh expansion}
\label {s:fourier}
For general background on Fourier--Walsh expansion, noise, and various applications in combinatorics, 
probability theory, and theoretical computer science see \cite {O'Do14,GaSt14}.
Let
$$\Omega_n = \{x=(x_1,x_2,\dots,x_n)
: x_i \in \{0,1\},i=1,2,\dots,n\}.$$ $\Omega_n$ is referred to as the discrete $n$-dimensional cube. (Elements in $\Omega_n$ are bitstrings of length $n$ of zeroes and ones.)
We consider real functions on $\Omega_n$ and define their Fourier--Walsh expansion.
The inner product for two real-valued functions $f$ and $g$ on $\Omega_n$ is defined by

$$\langle f,g \rangle=\sum_{x \in \Omega_n}2^{-n}f(x)g(x).$$
The 2-norm $\|f\|_2$ of $f$ is defined by $$\|f\|_2= \sqrt {\langle f,f \rangle }=\left(\sum_{x\in \Omega_n}2^{-n}f(x)^2 \right)^{1/2}.$$

We define now the Fourier--Walsh basis.
For $S\ne \emptyset \subset [n]$ define
$$W_S(x_1,x_2, \dots, x_n) = (-1)^{\sum \{x_i: i \in S \}},$$
and $W_\emptyset (x_1,x_2,\dots,x_n)=1$.
The $2^n$ real functions $W_S(x), S \subset [n]$ on $\Omega_n$
form an orthogonal basis for the vector
space of real functions on $\Omega_n$. (Here, $[n]$ stands for $\{1,2,\dots,n\}$.)

Let $f(x)$ be a real function  
on $\Omega_n$ (we will consider especially the case where $f$ describes a probability distribution on $\Omega_n$).
The Fourier--Walsh expansion of $f(x)$ is described by 
$$f(x)=\sum_{S:S \subset [n]} \widehat f(S)W_S(x),$$
where $$\widehat f(S)= \langle f,W_S \rangle.$$ 

{\bf Remark:} If we move from $0,1$-variables $x_1,\dots,x_n$ to $\pm 1$ variables
$u_1,\dots,u_n$ by letting $u_i = 1-2x_i$ then the Fourier--Walsh functions are simply the $2^n$
square-free monomials, namely $W_S(x_1,x_2,\dots, x_n)=\prod\{ u_i: i \in S\}$.

\subsubsection *{Parseval's formula and convolutions}
Parseval's formula asserts that $$\langle f,g \rangle=\sum_{S:S \subset [n]}\widehat f(S) \widehat g(S).$$
In particular,
$$\|f\|_2^2=\langle f,f \rangle=\sum_{S:S \subset [n]}\hat f^2(S).$$

For two functions $f$ and $g$ the convolution $f*g$ is defined by 
$$f*g(x)=\sum _{z \in \Omega_n}2^{-n}f(z)g(x+z).$$
(Here the sum $x+z$ is addition modulo 2.)
Another basic result about Fourier coefficients asserts that
$$\widehat {f*g} (S)= \widehat f(S)\widehat g(S).$$
In particular we get that
$$\sum _{S:S \subset [n]}\hat f^2(S) W_S= f*f.$$
Similarly, the multiplication of functions corresponds to convolutions for their Fourier coefficients.
If $h(x)=f(x)g(x)$, then
$$\widehat {h}(S) = \sum _{T:T \subset [n]}\widehat f(T)\widehat g(S \Delta T).$$
(Here, $\Delta$ is the symmetric difference of $S$ and $T$, which corresponds to sum modulo two of
the characteristic functions $1_S$ and $1_T$.)


\subsubsection *{The noise operator $T_{\rho}$}

Let $\rho, 0 \le \rho \le 1$ be fixed and let $x \in \Omega_n$. Consider the probability distribution $B_{\rho,x}(y)$ that assigns to the vector $y=(y_1,y_2,\dots, y_n)$ the probability 
$(1/2+\rho/2)^{n-k}(1/2-\rho/2)^k$, where $k=|\{i: y_i \ne x_i\}|$.     
In other words, the probability distribution $B_{\rho,x}$ describes the following random vector $y$: For the $i$th coordinate, with probability $\rho$, $y_i=x_i$, and with probability $1-\rho$, $y_i$ is a random bit, namely, with probability 1/2 it gets the value ``0" and with probability 1/2 it gets the value ``1". Moreover, these probabilities represent statistically independent events.     

For $\rho$, $0 \le \rho \le 1$ and a real function $f:\Omega_n \to \mathbb R$, define 

\begin {equation} 
\label {e:trho1}
T_\rho (f)(x)= \sum_{y \in \Omega_n} B_{\rho,x}(y)f(y).
\end {equation}
In words, $T_\rho(f)(x)$ is the expected value of $f(y)$ according to the probability distribution $B_{\rho,x}(y)$.

\begin {prop}
For a function $f:\Omega_n \to \mathbb R$, 
\begin {equation}
\label {e:trho2}
\widehat {T_\rho f} (S)= \rho^{|S|}\widehat f(S).
\end {equation}
\end {prop}


An easy way to prove it is to consider first $n=1$.
If $f(0)=a$ and $f(1)=b$ then
$$\widehat f(\emptyset ) = \frac {a+b}{2} ~~{\rm~~ and}~~ \widehat f(\{1\}) = \frac {a-b}{2}.$$
Now
$$T_\rho(f)(0)= \frac {1+\rho}{2} f(0) + \frac {1-\rho}{2} f(1) = \frac {1+\rho}{2} a + \frac {1-\rho}{2} b, $$ and similarly
$$T_\rho(f)(1)= \frac {1+\rho}{2} b + \frac {1-\rho}{2} a, $$
so $$\widehat {T_\rho (f)}(\emptyset)= \frac {T_\rho(f)(0)+T_\rho(f)(1)}{2}= \frac {a+b}{2} = \widehat f(\emptyset),$$
and 
$$\widehat {T_\rho(f)}(\{1\}) = \frac {T_\rho(f)(0)-T_\rho(f)(1)}{2} = 
\rho \frac{a-b}{2}
= \rho \widehat f(\{1\}).$$

When $n>1$ the effect of the noise is the tensor product of the
effects on the $k$-th bit, $k=1,2,\dots,n$, and this gives the formula
$\widehat {T_\rho (f)}(S)=\rho^{|S|}\widehat f(S).$

{\bf Remark:} The operator $T_{\rho}$ plays a central role in the analysis of Boolean functions
going back to the earlier works of Kahn, Kalai, and Linial \cite {KKL88} and Benjamini, Kalai, and Schramm \cite {BKS99}, see O'Donnell's book \cite {O'Do14}. 

\subsection {Random circuit sampling and the Google quantum supremacy experiment}

\subsubsection *{Random quantum circuits}

Every circuit $C$ with $n$ qubits 
describes a probability distribution ${\cal P}_C(x)$ for 0-1 vectors of length $n$. 
(In fact, it describes a $2^n$-dimensional vector of complex amplitudes; for every 0-1 vector
$x$, there is an associated amplitude $z(x)$ and ${\cal P}_C(x)=|z(x)|^2.$)
The quantum computer enables one to sample according to the probability distribution ${\cal P}_C(x)$,
with a considerable amount of noise. Let $m$ be the depth of the circuit $C$, namely the number of layers of computation. When $n$ and $m$ are not too large, classical computation enables one to compute the amplitudes themselves (and hence the probabilities ${\cal P}_C(x)$). Google's quantum supremacy claim was based on the fact that these classical simulations quickly become infeasible as $n$ and $m$ grow. 

When $C$ is a random circuit, the probability distribution ${\cal P}_C(x)$ behaves like a Porter--Thomas distribution; namely, the individual probabilities behave as if they have random statistically independent values drawn from the exponential distribution (that are then normalized).
Of course, an instance of a Porter--Thomas distribution depends on $2^n$ random variables, 
while the $2^n$ values of ${\cal P}_C(x)$ depend on the circuit $C$ whose description requires a polynomial number of parameters in $n$. 

\subsubsection* {The Google noise model}

The Google basic noise model is 
\begin{equation}
\label{e:gnm}
{\bf N}_C(x) =  \phi {\cal P}_C(x) + (1-\phi) 2^{-n},
\end{equation}
where $\phi$ is the {\it fidelity}, a parameter that roughly describes the quality of the sample. Roughly speaking, Google's noise model assumes that if all components of the quantum computer operate without errors then a bitstring will be drawn according to ${\cal P}_C(x)$, and if an error occurs then the quantum computer will produce a random uniformly distributed bitstring. 

\subsubsection *{The linear cross-entropy estimator}

Based on their noise model (and the fact that the distribution ${\cal P}_C$ is an instance of a Porter--Thomas distribution), the Google paper \cite {Aru+19} describes a statistic called the {\it linear cross-entropy estimator}, 
denoted by ${\cal F}_{XEB}$ (and sometimes simply by $XEB$).
Once the quantum computer produces a sequence ${\bf \tilde x}$ of $N$ 
independent bitstrings ${\bf \tilde x}= ({\bf \tilde x}^{(1)},{\bf \tilde x}^{(2)},\dots, {\bf \tilde x}^{(N)})$, the estimator ${\cal F}_{XEB}$ of the fidelity is computed as follows:

\begin{equation}\label{e:fxeb}
{\cal F}_{XEB}({\bf \tilde x})=\frac{1}{N}\sum _{i=1}^N 2^n{\cal P}_C({\bf \tilde x}^{(i)}) -1.
\end {equation}

In Section 6 of \cite {RSK22} we described a refined noise model 
based on the effect of readout errors, and 
combining the Google statistical method with our improvements allows estimating the effect of 
readout errors from the data. (In \cite {RSK22} this was carried out for $n=12,14$.) 

The Google quantum supremacy claim is also based on an a priori prediction of the fidelity of a circuit based on the probabilities of error for the individual components (Formula (77) in the supplement to \cite {Aru+19}):

\begin {equation}
\label {e:77}
~\widehat \phi~=~ \prod_{g \in {\cal G}_1} (1-e_g) \prod_{g \in {\cal G}_2} (1-e_g) \prod_{q \in {\cal Q}} (1-e_q).
\end {equation}

Here ${\cal G}_1$ is the set of 1-gates, ${\cal G}_2$ is 
the set of 2-gates, and
${\cal Q}$ is the set of qubits. For a gate $g$, the term $e_g$ in the 
formula refers to the probability of an error 
of the individual gate $g$. 
For a qubit $q$, $e_q$ is the probability of a readout error when we measure the qubit $q$. 
The average values 
of 1-gate errors is 
$0.0016$,  of 2-gate errors 
is 
$0.0062$, and the average value of readout errors is $0.038$. 
We refer to equation \eqref{e:77} as ``Formula (77)" throughout the paper.

{\bf Remark:}  For large values of $n$ and complicated 
circuits $C$ computing the value of ${\cal P}_C(x)$ is infeasible 
and the argument of \cite {Aru+19} and other quantum supremacy experiments is based on extrapolation.
The ${\cal F}_{XEB}$ estimator is used to verify the prediction of Formula (77) when either $n$ is small ($n \le 42$) or the circuit $C$ is simple (and allows efficient computation of ${\cal P}_C$). From this \cite {Aru+19} infers that the prediction of Formula (77) continues to hold when the circuit $C$ is complicated and $n$ is large. 

\subsection {A brief review of fidelity estimators studied in \cite {RSK22}}

In \cite {RSK22} we made a comprehensive study of fidelity estimators for samples of random quantum  circuits. We considered the following estimators.  
\begin {enumerate}
\item
Google's main ${\cal F}_{XEB}$ estimator referred in \cite {RSK22} as ``$U$" (see \eqref{e:fxeb}). 
\item 
An improved unbiased version of ${\cal F}_{XEB}$ estimator referred in \cite {RSK22} as ``$V$". (See \eqref{eq:V}, Section \ref {app:est}.)
\item 
MLE (maximum likelihood estimator). (See \eqref{eq:MLE}, Section \ref {app:est}.)
\item 
$T$ - an estimator based on the numbers of repeated bitstrings in the data. $T$ is closely related to  ``SPB" (``speckle purity benchmarking") in the supplement of \cite {Aru+19}. 
See \eqref{eq:T_squared}, Section \ref{app:est}.

\item A priori prediction of the fidelity based on the fidelities of individual components of the circuits via Formula (77) (equation \eqref {e:77}). 

\item  Estimators for $\phi_{g}$, the probability of ``no gate errors", and for $\phi_{ro}=\phi_g-\phi$, the probability of no gate errors and at least one readout error. Estimators for $q$, the rate of readout error. 

\end {enumerate}

The formulas for MLE, $V$, $T$, a refined version of $T$ called ``$S$," are given in Section \ref{app:est} of the appendix. 
Several estimators for $\phi_{ro}$ and for the parameters $s$ and $q$ 
in formula \eqref {e:noise-model-rho} and the connection to Fourier--Walsh coefficients are described in Section \ref{s:fourier-readout} below.

Generally speaking, the estimators ${\cal F}_{XEB}$, $V$ and $MLE$ are based on the correlation between the data of bitstrings and the ideal probabilities ${\cal P}_C(x)$. Regarding the probability distribution ${\cal P}_C(x)$ as the ``primary signal" that we want to detect in the data, these estimators give a measure for the strength of this primary signal in the data. 

The estimators for $\phi_{ro}$ in item 6. above are based on detecting a ``secondary signal" in the data, defined by Equation (6.1) in \cite {RSK22} and denoted there by ${\cal N}_C^{ro}(x)$. The secondary signal can be written in terms of Fourier--Walsh coefficients 
as follows:
\begin{equation}
\label{e:secondary}
{\cal N}_C^{ro}(x)=T_{(1-2q)}({\cal P}_C(x))- (1-2q)^n {\cal P}_C(x),
\end{equation}

\noindent
where $T_{(1-2q)}$ is the operator described by \eqref{e:trho1} (and \eqref{e:trho2}). The secondary signal ${\cal N}_C^{ro}(x)$ depends on the circuit $C$, and it is (almost) orthogonal to the primary signal ${\cal P}_C(x)$. 
Our estimators for $\phi_{ro}$ are closely related to the Fourier--Walsh coefficients of the samples and we discuss the relation in Section \ref {s:fourier-readout} below. 

\comment { Here is a brief review of our previous papers. In our first paper \cite {RSK22} we mainly studied fidelity estimators and made a preliminary comparison of the empirical distribution of the samples and that of the Google noise model. The Fourier methods described here are related to fidelity estimators based on the readout errors from Section 6 of \cite {RSK22}. (As a matter of fact, we realized here that a certain statistical assumption of independence for the estimators of Section 6 of \cite {RSK22} is incorrect and analyze the consequences.)   
Our second paper \cite {KRS22d} provided a detailed description of data and information regarding the Google experiment and listed some confirmations, refutations, weaknesses, and concerns. 
Our third paper \cite {KRS23} presented statistical analysis of various aspects of the Google experiment. On the technical side, this paper represents noisy simulations based on Google's and IBM's simulators that were not applied in our earlier papers
}

\newpage
\section {Randon circuit sampling - Fourier--Walsh expansion and noise} 
\label {s:rcs-four}
\subsubsection* {Fourier--Walsh coefficients of a Porter--Thomas distribution} 

\begin{figure}
\centering
\includegraphics[width=\textwidth]{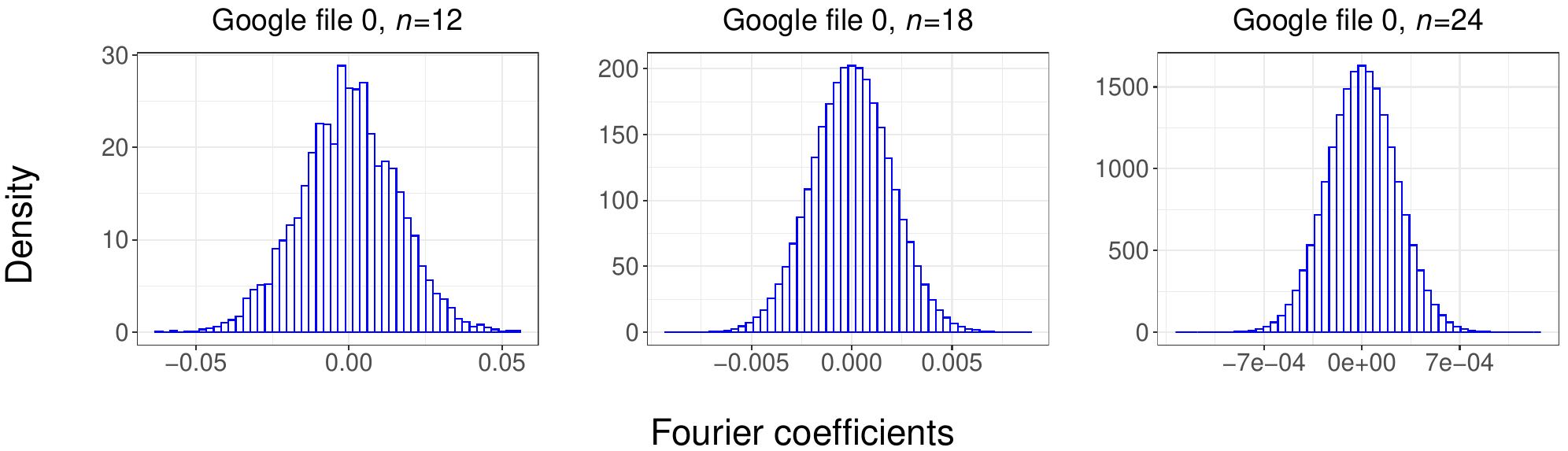}
\caption{A histogram of the Fourier--Walsh coefficients $\widehat P(S)$ for different Google files.} 
\label{fig:Histogram_Fourier_coefficients}
\end{figure}

Let $P(x)={\cal P}_C(x)$ be the probability distribution described by Google's random
circuit $C$, or, in fact, any other Porter--Thomas distribution. 
We can expect that when $S \ne \emptyset$,
$2^n \widehat P(S)$ behaves like a random  alternating sum of $2^n$ exponential random variables 
with variance 1.
Therefore, the individual Fourier--Walsh coefficients $\widehat P(S)$
will be very close to Gaussian, as we can observe in Figure \ref{fig:Histogram_Fourier_coefficients}.

\subsubsection *{Fourier--Walsh description of Google's noise model}

We use the letter $N$ to describe a noisy distribution
based on $P$. Let $$N (x)=\phi P(x)+ (1-\phi)2^{-n}$$
denote the distribution described by the Google noise model.
The Google model has a simple effect on the Fourier
coefficients
$\widehat N(\emptyset)=\widehat P(\emptyset)=2^{-n}$, and
$$\widehat N(S)=\phi \widehat P(S), {\rm ~~ if ~~} S \ne \emptyset.$$ 

\begin{figure}
\centering
\includegraphics[width=\textwidth]{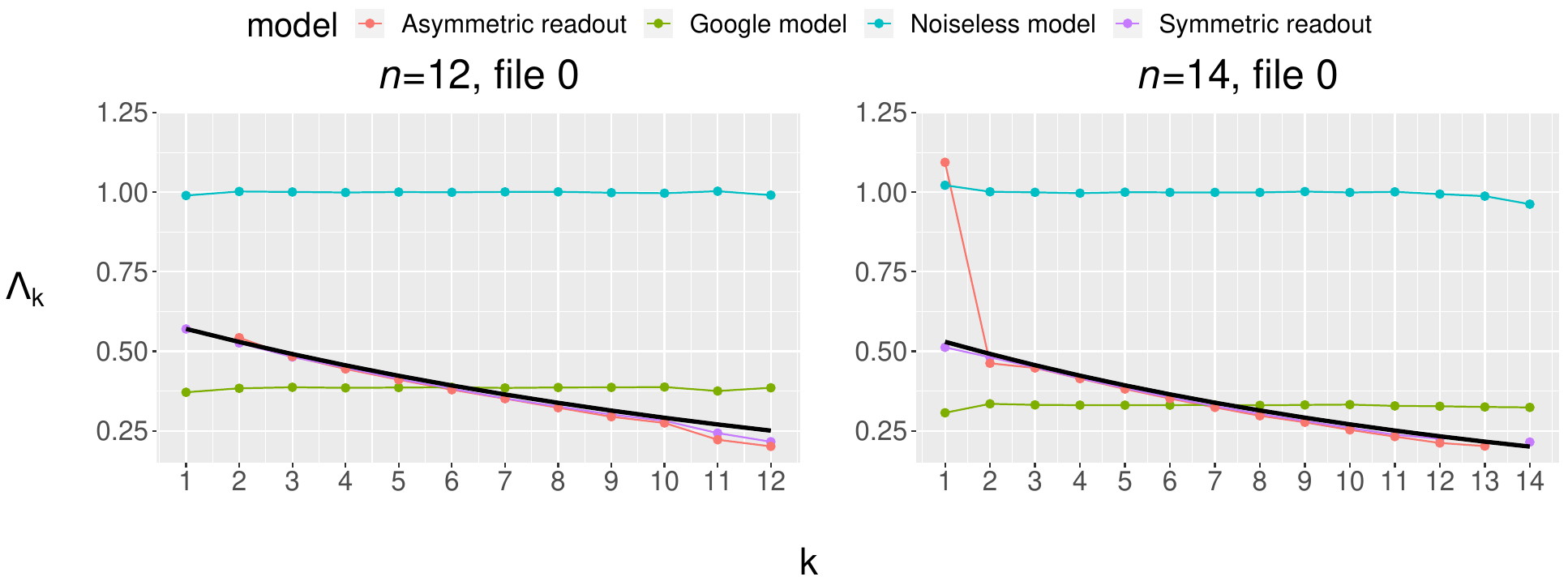}
\caption{Simulation of the effect of various noise models on the degree $k$ Fourier-Walsh contributions $\Lambda_k$ as defined in equation \eqref{e:Lambda}. The noiseless model is the Google noise model, with $\phi=1$. For all the other models we took $\phi=0.3862$ when $n=12$, and $\phi=0.332$ when $n=14$. The Google model is given in \eqref{e:gnm}. The symmetric readout model is given in \eqref{e:secondary} with $q=0.038$. The asymmetric readout model is detailed in \cite{RSK22}, and in this simulation we took the probability $q_1$ that 1 is read as 0 is 0.055 and the probability $q_2$ that 0 is read as 1 is  0.023. The simulation is based on sampling $N=500,000$ samples from the corresponding model, where the base probabilities are the first Google file, with $n=12$ or $n=14$. The solid black line is given by Equation \eqref{eq:reference_readout}. }
\label{fig:1}
\end{figure}

\subsubsection *{Symmetric readout noise: exponential decay of the Fourier--Walsh coefficients}

For a distribution $P$ consider next a readout noise operator  $N$ where  the probability that a bit
is flipped is $q$, and assume that there are no other errors. 
$N(x)$ is the probability of observing $x$ when bitstrings are generated by the probability distribution $P$, and if $x \oplus y$ is generated we observe $x$ if there is a flip for each $i$ with $y_i=1$. 
As stated in Section \ref {s:fourier} $N(x)$ with the parameter $q$ is the same as the noise operator $T_\rho$ with $\rho=(1-2q)$,  and Equation  \eqref {e:trho2} asserts that 


$$
  \widehat N(S) = (1-2q)^{|S|}\widehat P(S).
  $$
\comment {
An easy way to prove it is to consider first $n=1$.
If $P(0)=a$ and $P(1)=b$ then
$\widehat f(\emptyset ) = (a+b)/2$ and $\widehat P(\{1\}) = (a-b)/2$.
Now
$N(0)= (1-q)P(0) + q P(1) = (1-q)a+tb $ and $N(1)= (1-q)P(1) + q f(0) = (1-q)b+qa$
So $$\widehat N(\emptyset)=a+b$$
(as before) but $$\widehat N(\{1\}) = ((1-q)a+tb)-((1-q)b+ta))/2 = (1-2q)(a-b)/2
= (1-2q) \widehat P(\{1\}).$$

When $n>1$ the effect of the noise is the tensor product of the
effect on the $k$-th bit, $k=1,2,\dots,n$. And this gives the formula
$$\widehat N(S)=(1-2q)^{|S|}\widehat P(S).$$
}
{\bf Remark:} The symmetric readout noise (denoted there by ${\cal N}_C^{ro}$) of \cite {RSK22}
is of the form 
\begin{equation} 
\label {e:symm-noise-gates}
\phi_g T_{(1-2q)}({\cal P}_C(x)) + (1-\phi_g) \frac {1}{M}
\end{equation} 
where $\phi_g$ is the probability of no gate errors. 
This model is based on the assumption that any gate error causes the distribution to be uniform. 
It follows that the effect of the symmetric noise model 
on the Fourier--Walsh coefficients 
is that the $\widehat P (S)$ is
multiplied for $S \ne \emptyset$ by
\begin{equation}
\label{eq:reference_readout}
    \phi_g \cdot(1-2 \cdot 0.038)^{|S|}.
\end{equation}


\subsubsection* {Asymmetric readout errors} 

We expect
that for any type of realistic noise, Fourier coefficients $\widehat P(S)$ will be reduced
exponentially with $|S|$. 
In particular, we expect an exponential decay with the
same constant $\phi_g \cdot (1-2 \cdot 0.038)^k$
for the asymmetric readout errors
from \cite {RSK22} based on  
the actual readout errors of the Google device: the probability $q_1$ that 1 is read as 0 is 0.055 and the probability $q_2$ that 0 is read as 1 is  0.023.
This is indeed supported by simulations (see, Figure \ref{fig:1}).
(We can also expect the same behavior to hold if we base the
error model on the precise readout errors which are somewhat different for different qubits.)



\subsubsection* {The effect of gate errors}

In Section 6 of \cite {RSK22} for our readout noise models, 
we assumed 
that the effect of a gate-error is that of changing
the distribution to a uniform distribution. 
This assumption means that the effect of gate errors is to multiply all
nonempty Fourier coefficients by the same constant. Here we revisit this assumption, and studying the effect of gates errors, mainly by simulation, will play a central role in Section \ref {s:gates} below. As it turns out, gate errors, similar to readout errors (but to a smaller extent), have a larger effect on high-degree Fourier coefficients. 
We note that such an effect of gate errors is expected also from the findings of \cite  {AGLLV22} and \cite {Gao+21}.

\subsubsection*{Further remarks regarding noise modeling}

A schematic picture regarding the effect of noise is as follows. Starting with the ideal distribution of bitstrings we first consider the effect of gate errors and on the resulting probability distribution we apply the effect of readout errors. Two remarks: first, a full treatment of noise needs to be in terms of the actual quantum process rather than in terms of the effect on the samples. Second, a finding from Section 4.7 of \cite {RSK22} is that the effect of noise on the variance of the empirical distribution is considerably larger than what can be expected by simple forms of readout errors and gate errors. (This is manifested by the large values of estimators ``$T$" and ``$S$" for the fidelity.) This additional form of noise is related to the notion of ``SPB" (``speckle purity benchmarking") from the Google paper (Section VI.C.4 of the supplement of \cite{Aru+19}).

\newpage
\section  {Statistical tools for Fourier analysis  }
In this section we refine statistical tools from \cite {Aru+19,RSK22} based on Fourier--Walsh expansion. 

\label {s:stat-fourier}
\subsection {Estimating the effect on degree $k$ Fourier--Walsh coefficients}

Our first task is to estimate from the experimental bitstrings
the effect of the noise on the Fourier coefficients of level $k$. As before let $P$ denote a probability distribution and we write 
$$P_k= \sum_{S:|S|=k} \widehat P(S)W_S.$$
Note that $P=P_0+P_1+P_2+\dots +P_n,$ and
that $P_0$ is the uniform probability distribution. 
We will try to estimate the effect of noise on the different Fourier levels, namely to describe the empirical (noisy) distribution $A(x)$ of the 
number of samples that gives $x$ as 
\begin {equation}
\label{e:noisy-fourier}
A(x) = \alpha_0P_0 + \alpha_1 P_1 +\dots + \alpha_n P_n,
\end {equation}
and to empirically estimate
the various $\alpha_k$, $k=1,2,\dots,n$. (We take $\alpha_0=N$.)

Our second task will be to use Fourier tools to compute fidelity estimates based on readout errors proposed in Section 6 of \cite {RSK22}. 
\subsection {Estimating the $\alpha_k$s and the connection with fidelity estimators $U$ and $V$}
Let $P(x)$ be a probability distribution on bitstrings of length $n$. (Recall that $M=2^n$.) Define 

\begin {equation}
  \gamma_k = \langle P,P_k \rangle = \|P_k\|_2^2.
\end {equation}

Our statistical task is the following: we assume that we are given a sample 
$z_1,z_2,\dots ,z_N$ of a noisy version of $P$ described by equation 
\eqref {e:noisy-fourier} and we wish to estimate the values of $\alpha_1,\alpha_2,\dots, \alpha_n$. Define
  
  \begin {equation}
    U_k = \sum _{i=1}^N P_k(z_i).
\end {equation}

When $P={\cal P}_C$, the probability distribution described by a random circuit,
$$U_1+U_2+ \dots + U_n=   \sum _{i=1}^N P(z_i).$$
It follows that up to normalization $U_1,U_2,\dots, U_n$ refines Google's main fidelity estimator ${\cal F}_{XEB}$, namely 
\begin {equation} {\cal F}_{XEB}=  \frac {M}{N} (U_1+U_2+ \cdots + U_n),
\end {equation}

 \comment {Over all Porter Thomas distributions the expected value of $<P,P_k>$ will be
 proportional to ${{n} \choose {k}}$ since all Fourier
 coefficients $\widehat P(S)$, $S \ne \emptyset$
 behave in the same way.} 
 
 For a specific Porter--Thomas $P={\cal P}_C$ and a  sample $z_1,z_2,\dots, z_N$ 
 described by equation \eqref {e:noisy-fourier},
define 
 
 \begin {equation} 
 \label {e:Lambda}
 \Lambda_k=\frac {U_k}{\gamma_k}.
 \end {equation} 
 $\Lambda_k$ is an estimator for $\alpha_k$, namely it estimates the effect of the noise on the level-$k$ Fourier--Walsh coefficient. 


{\bf Remark:} We note that to compute $\widehat P(S)$ and the functions $P_k$, $k=0,1,\dots ,n$, we need
the amplitudes (giving the values of ${\cal P}_C(x)$) for all bitstrings $x$
and not only for those in the samples.





\subsection {Fourier description of $\Lambda_k$ and implementation}




Let $A(x)$ be the number of samples that gives $x$.
Now $U$ is $\frac {M}{N} \cdot \langle A,P \rangle$, and since we can compute
the Walsh--Fourier expansion of $P$ and of $A$ we get
$$M \cdot U_k = \frac {M}{N} \sum _{S:|S|=k} \widehat A(S) \widehat P(S).$$
By the same token, $$\gamma_k = \langle P,P_k \rangle = \sum _{S:|S|=k} \widehat P^2(S).$$
(To see this, note that $\langle P_k,P \rangle = \langle P_k,P_k \rangle $ since $\langle P_k, W_S \rangle =0$ if $|S| \ne k$.)

We implemented this computation for the Google data and various simulations for the range $12 \le n \le 28$, and this allowed us to 
compare the empirical behavior of the degree $k$ contribution and the theoretically expected one. 

Let us draw attention to the case $k=n$. We obtain that 
\begin {equation}
\Lambda_n = \frac {1}{N} \cdot \frac {\widehat A([n])}{\widehat P([n])},
\end {equation}
in other words, $\Lambda_n$ is the ratio between the top Fourier-Walsh coefficient of the empirical probability and that of the ideal probability. The coefficient $\widehat A([n])$ is simply the number of bitstrings with even number of ones minus the number of bitstrings with odd number of ones. 

{\bf Remark:} Our first implementation was based on computing $\widehat P(S)$, and then
the functions $P_k$, $k=0,1,\dots n$. We then computed $U_k$ by \nopagebreak
$U_k = 
\sum P_k(z_i)$. Since there are quick algorithms for
the Fourier--Walsh transform,
the part which was computationally costly was the computation of the functions $P_1,P_2,\dots,P_n$.
As it turned out
computing the functions $P_k$ is not needed. 



\subsection {Fourier description of fidelity estimators based on symmetric readout errors}
\label{s:fourier-readout}

In this subsection we discuss some estimators considered in Section 6 of \cite {RSK22} that can be described and efficiently computed using Fourier--Walsh transform. A general schematic model for the noisy signal from \cite {RSK22} is of the form
\begin {equation}
\phi {\cal P}_C + (\phi_g-\phi){\cal N}_C^{ro} + (1-\phi_g){\cal N}_C^g
\end {equation} 
The first term describes the noiseless distribution, the second term describes the
effect of readout errors given that there are no gate-errors and at least one readout error, and the third term described the event that there are 
gate errors. 
In Section 6 of \cite {RSK22} we defined $\phi_{ro}=\phi_g-\phi$ and proposed statistical ways to estimate $\phi_{ro}$ and $\phi_g$. Here we note that we can use Fourier--Walsh  transform to speed up computations of these estimators. 
We also note that an estimator for $\phi_{ro}$  gives alternative ways to estimate $\phi$ based on a ``secondary'' signal of the Google samples and Formula (77) as follows:

\begin {equation}
\label {e:alt-phi}
\text{alt-}\phi \approx (\phi_g-\phi)\frac {1}{(1-q)^{-n} - 1} = \phi_{ro}\frac {1}{(1-q)^{-n} - 1} .
\end {equation}


\subsubsection {${\cal F}_{XEB}$-style estimator for $\phi_{ro}$ }

The symmetric readout error from Section 6 of \cite {RSK22} is 

\begin {equation}
\label {e:symm-readout}
\phi {\cal P}_C + (\phi_g-\phi){\cal N}_C^{ro} + (1-\phi_g)/M = \phi_g T_{(1-2q)} ({\cal P}_C) +(1-\phi_g)/M,
\end {equation}
and one of our statistical estimators was based on computing
\begin{equation}\begin{split}\label {e:V-phiro} 
\frac {1}{N} \sum_{i=1}^N {\cal N}_C^{ro}(x_i) 
&= \langle A(x),{\cal N}_C^{ro}(x) \rangle \\
&=\langle A(x), (T_{(1-2q)}{\cal P}_C(x)-(1-q)^n{\cal P}_C(x)) \rangle,
\end{split}\end{equation}
  where $q=0.038$.
  This formula gives us a quick way to compute the estimates for $\phi_{g}$ (and hence to estimate $\phi_{ro}=\phi_g-\phi$ 
  without the need to compute the distribution ${\cal N}_C^{ro}(x)$ itself. The estimator for $\phi_{ro}$ described in Equation (6.6) of \cite {RSK22} is nearly unbiased with respect to the samples and the random circuits, in the same sense that ${\cal F}_{XEB}$ is unbiased. (It is possible to improve this estimate to a $V$-style estimator for $\phi_{ro}$ but we did not implement this version.)

\subsubsection {An MLE estimator for $\phi_{ro}$}
\label {s:mle-phiro}
 Equation \eqref {e:V-phiro} describes an estimator for $\phi_{ro}$ based on correlation with the secondary signal which is analogous to the estimator ${\cal F}_{XEB}$ for the fidelity that is based on the correlation with the primary signal ${\cal P}_C(x)$. In \cite {RSK22} we gave also an MLE estimator for $\phi_{ro}$, and for the pair of parameters $(\phi,\phi_{ro})$ which are unbiased even for every specific circuit, and also for these estimator, Fourier calculations make the computations considerably more efficient.  
 When it works,  the MLE $\phi_{ro}$ estimator appears to have smaller variance  (see Table \ref {t:phi-rho-comp} for the experimental data and compare Tables \ref {t:mle-phiro-sim} and \ref {t:QVM} for the QVM-simulation data) and this was the case for the Google samples, both from the experiment and from simulations. In some other cases, however, (like the IBM data) the optimization algorithm for the MLE estimator for $\phi_{ro}$ did not converge to a global minimum. It is possible to describe also a $V$-style estimator for $\phi_{ro}$ which is a normalized improved version of Equation (6.6) of \cite {RSK22} which will be unbiased even for a single circuit and will not require optimization. 

\subsubsection {An MLE estimator for $s$ and $q$}
 
Another (closely related) statistical method from Section 6 of \cite {RSK22} was using MLE to give 
the best fit of the empirical data with the two-parameter noise model given by Equation \eqref {e:noise-model-rho}, namely by
$$
sT_{(1-2q)} ({\cal P}_C) + (1-s)/M.
$$

Also for this case, fast Fourier--Walsh transform allows efficient computation. We expect that this two-parameter noise model and the statistical estimation of the parameters  is relevant to the study of samples coming from NISQ computers beyond the specific case of the Google quantum supremacy experiment. See Section \ref {s:sq-simulations} in the appendix for some data. 

\subsubsection {Statistical assumptions} 
  
The estimators for $\phi_{ro}$ from Section 6 of \cite {RSK22} were based on two assumptions, first that the effect of readout errors ${\cal N}_C^{ro}(x)$ is uncorrelated to the main signal ${\cal P}_C(x)$, and second that ${\cal N}_C^{ro}$ is uncorrelated with 
$N_g$. It turned out the second assumption is incorrect and we discuss it in Section 
\ref {s:gates}. As we already mentioned under these statistical assumptions in our basic noise model given by Equation \eqref {e:noise-model-rho}, $q$ is the rate of readout errors and $s=\phi_g$.  


\subsubsection {Estimating the parameters for an asymmetric readout model}
  
A more accurate model (while fairly close) for the noise takes into account the different probabilities, 
$q_1$ that 1 is read as 0 and $q_2$ that 0 is read as 1. 
For the Google experiment these probabilities are substantially different, $q_1=0.055$ and $q_2=0.023$. 
 In Section 6 of \cite {RSK22} we considered MLE estimators for the triple ($\phi_g$, $q_1$, $q_2$). Also in this case the effect of the noise can be described as a convolution and our Fourier methods could be computationally useful for computing this estimator. 

\newpage
\section {Fourier analysis of data from the Google experiment and from simulations}
\label {s:gates}
\subsection {The Fourier analysis of the Google data}  
\label {s:google}

\begin{figure}
\centering
\includegraphics[width=\textwidth]{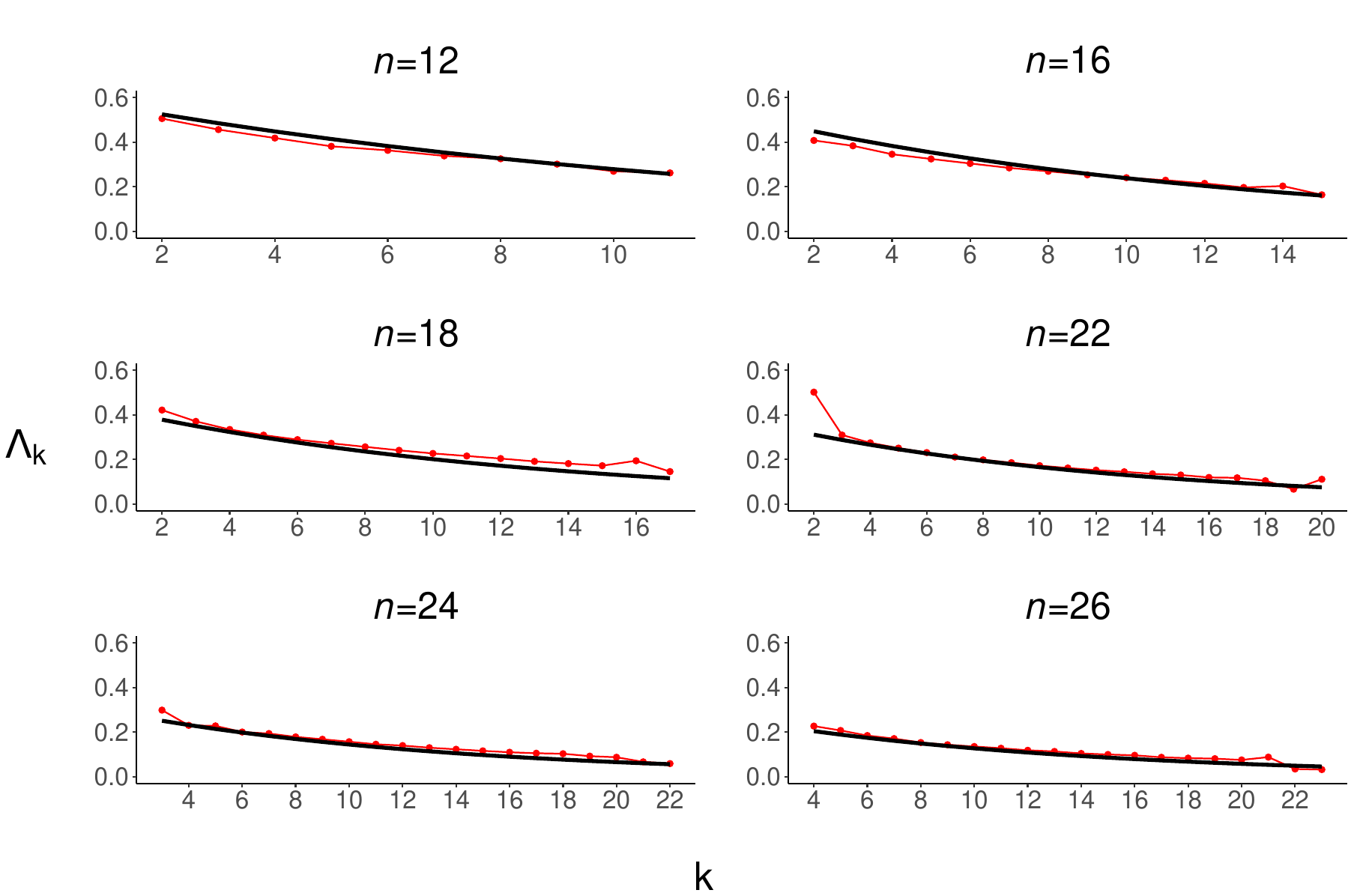}
\caption{The decay of degree $k$ Fourier-Walsh contribution coefficients averaged over the ten experimental (full) circuits for the Google experimental data for $n=12,16,18,22,24,26$. The degree-$k$ Fourier contributions $\Lambda_k$ are defined in equation \eqref{e:Lambda}. The solid black line is based on the symmetric readout noise model given by Equation \eqref {eq:reference_readout}, with $\phi=0.3862, 0.2828,0.2207, 0.1554,0.1256,0.1024$ for $n=12,16,18,22,24,26.$ Extreme values of $k$ are unstable. 
}
\label{fig:Fourier_Delta_Google_data}
\end{figure}

We compared the decay of degree $k$ Fourier coefficients for the Google data to that of ``pure" (symmetric) readout errors. Figure \ref{fig:Fourier_Delta_Google_data} describes the average contributions according to the Fourier degrees for the Google data of the ten circuits with $n=12,16,18,22,24,26$ qubits, compared to the estimated contribution based on readout errors. For extreme values of $k$ the estimate $\Lambda_k$ is based on very few observations, and thus, very unstable. We omitted extreme values of $\Lambda_k$ from the figure (and subsequent figures) to keep its scale. See Section \ref {s:stab} for the full data and discussion.


Overall, the behavior of the Fourier contribution according to Fourier degrees for the Google samples exhibits a decay which is similar 
to what could be expected from the effect of readout errors with no additional effect of gate errors. 

\begin{figure}[h]
\centering
\includegraphics[width=\textwidth]{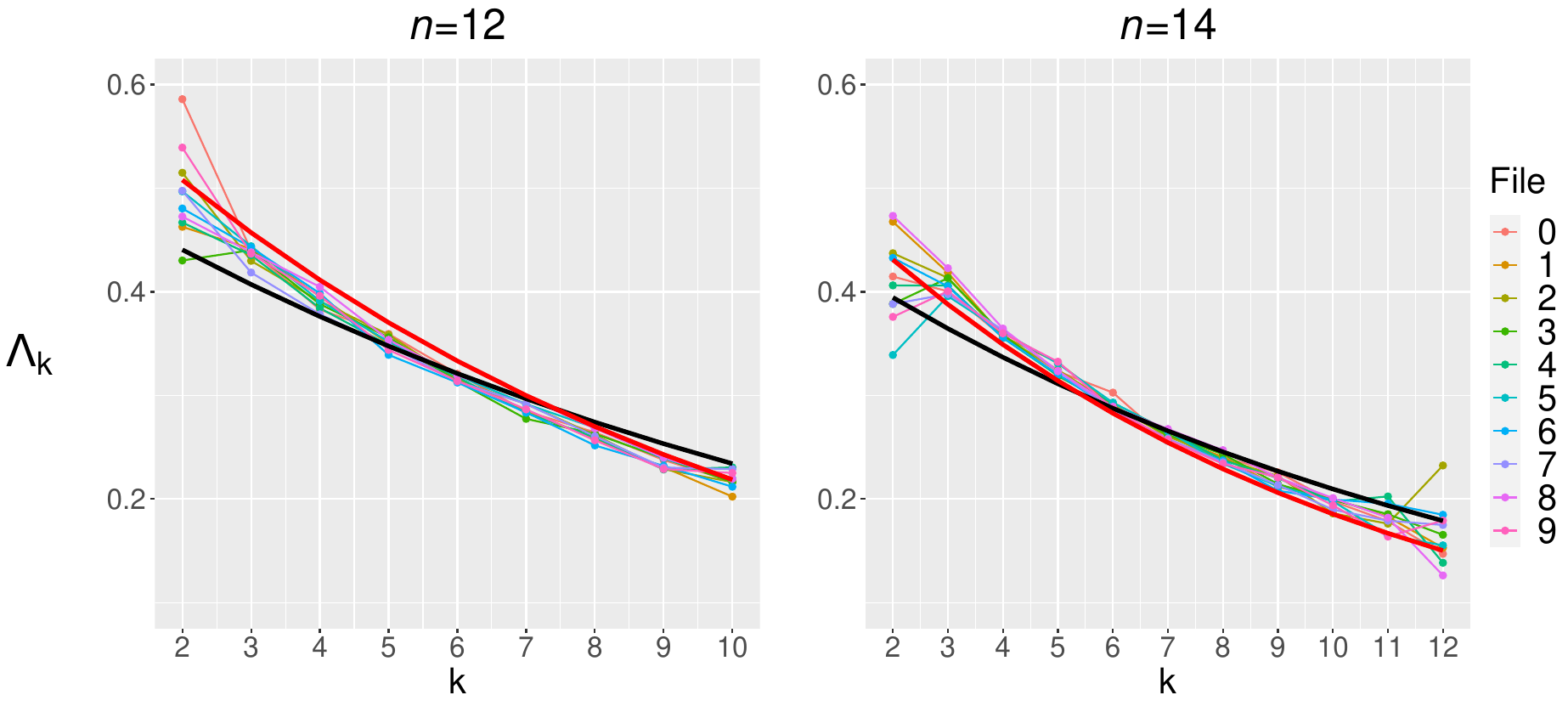}
\caption{The decay of degree $k$ Fourier coefficients for simulation with Google’s full noise model of the Google circuits for 12 and 14 qubits. The main decrease in the Fourier contribution is based on the readout errors and it appears that the gate errors contribute an additional effect of a similar nature. The solid black line describes the theoretical effect of readout errors \eqref{eq:reference_readout} under the assumption that every gate error leads to a uniform random bitstring. (It is based on $q=0.038$, $\phi_g=0.516$ for $n=12$ and $\phi_g= 0.462$ for $n=14$. The values were derived from Table \ref{t:QVM} using Equation \refeq{eq:phi_ro}). The solid red line is based on the best fit: For $n=12$, $q=0.053$, and $s=0.627$, for $n=14$, $q=0.047$ and $s=0.532$ and. Extreme values of $k$,  are unstable and are omitted. See Section \ref {s:stab} for the full data and discussion. }
\label{fig:ohad12+14}
\end{figure}

\begin{figure}[h]
\centering
\includegraphics[width=\textwidth]{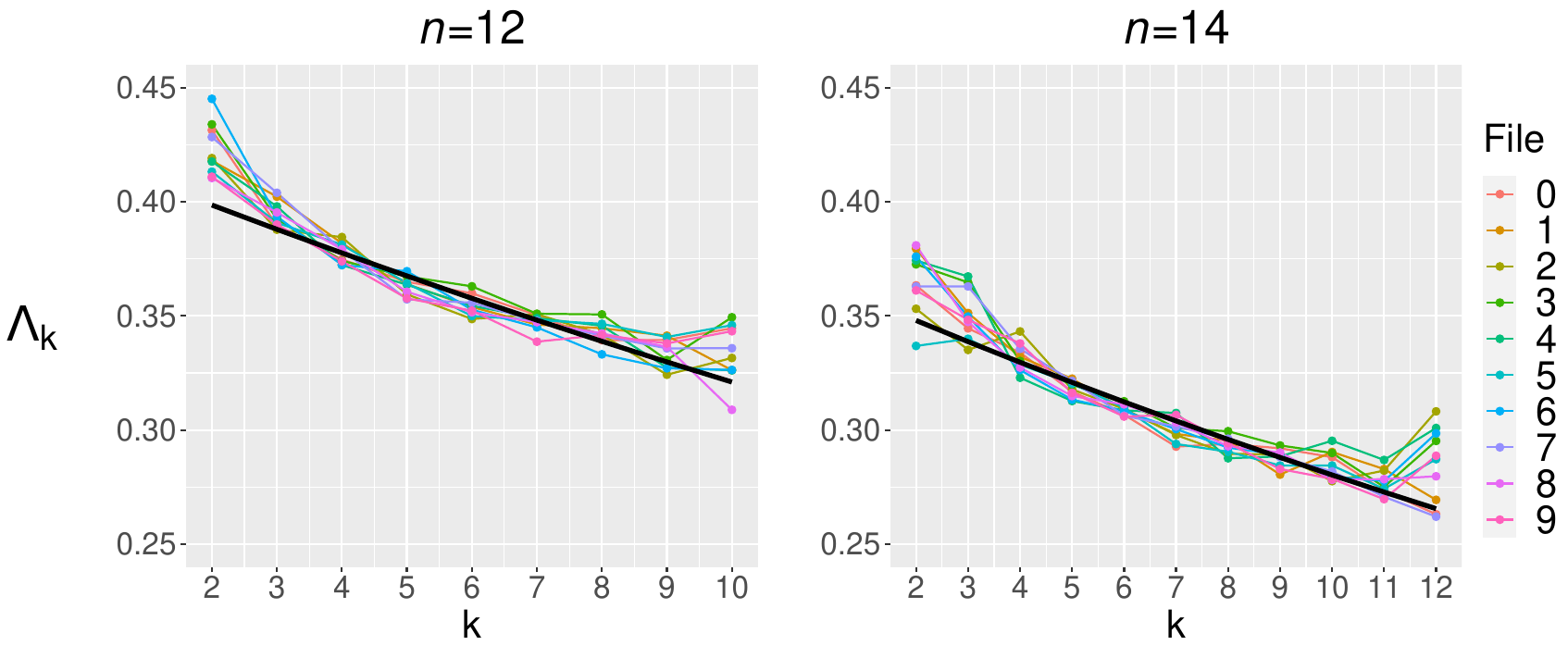}
\caption{The effect of gate errors on the degree $k$ Fourier-Walsh contribution $\Lambda_k$. This computation is based on simulations with Cirq of the Google files with 12 qubits with 1-gate depolarization errors on all 1-gates (including those for the calibration). The error rate is adjusted so that the overall fidelity is similar to the Google experimental fidelity.  For these simulations the best fit for the parameters $s$ and $q$ in the model given by Equation \eqref {e:noise-model-rho} was (averaged on 10 simulation files) 
$s=0.4207$ and $q=0.013$, for $n=14$ we have $s=0.3673$, $q=0.014$.
(Extreme values of $k$,  are unstable and are omitted.) 
} 
\label{fig:gates1}
\end{figure}

\subsubsection*{Non-stationary behavior and Fourier}

We studied non-stationary behavior in Section 3.5.1 of \cite {KRS23} by comparing the distance between the empirical distributions of the first and second halves of samples to the distance between random partitions of the samples into two halves, and also by considering drifting in the behaviors of individual qubits. We refined this study and considered such a comparison (for $\ell_2$ distances) according to the Fourier degrees. Our preliminary study
suggests that the non-stationary behavior for bitstrings of length 12 in the Google data is primarily caused from drifts in the behavior of a few individual qubits and pairs of qubits.
(This would allow, in this case, to predict future behavior from past data.)

\subsection {Simulations}
\subsubsection* {Google's and IBM's simulators}
Simulations suggest that noise described by depolarizing gate errors cause a larger reduction for the contribution of higher Fourier--Walsh degrees 
and thus have similar behavior to that of $T_\rho$ for some value of $\rho$. We collected data from three types of simulations.  The first type of simulation ``Weber QVM simulations with the full noise model" \cite {Weber} is the closest simulation available to us for the Google Sycamore experiment; it is based on simulation of the Google circuits with Google's complete noise model.\footnote{Our choice of parameters in applying the Weber QVM simulator was based on averaged (symmetric) readout errors and gate errors.} Here, Figure \ref{fig:ohad12+14} describes the Fourier--Walsh behavior for simulations based on the ten circuits of the Google quantum supremacy experiment. The main decrease in the Fourier contribution is caused by the readout errors and it appears that the gate errors contribute an additional effect of a similar nature. Quantitatively, for these simulations the effective readout error (based on MLE estimation of the parameters $s$ and $q$) was 0.053 for 12-qubit simulations and 0.047 for 14-qubit simulations compared to the physical readout error of 0.038. 

The second type of simulation was a 12-qubit simulation of the Google circuits where  1-gate depolarization errors on all 1-gates (including those for the calibration) were added. 
The Fourier--Walsh behavior is described in Figure \ref {fig:gates1} where we see that gate errors have a larger effect on higher degree Fourier--Walsh coefficients. (The rate of error was chosen to yield a similar value for the fidelity to that of the experiment.) Quantitatively, while the simulation had no readout errors the effective readout error based on MLE estimation of the $(s,q)$ pair was 0.014.


The third type of simulation was the IBM simulation program ``Fake Guadalupe” based on  IBM's 16-qubit quantum computer ``Guadalupe," see \cite {Guad}. We implemented random circuits (created by the Cirq simulator itself) with 12 
qubits. (These circuits are very different from Google's random circuits.) 
The values of $\Lambda_k$ for ten circuits are described in Figure \ref {fig:Fourier-fake-guadalupe} in Section \ref {s:ad} of the appendix. In this case the average readout error is 0.022 and the effective readout error is 0.66. In Figures \ref{fig:ohad12+14}, Figure \ref {fig:gates1}, and Figure \ref {fig:Fourier-fake-guadalupe} we omitted the values $k=1,n-1,n$. In these  cases the estimate $\Lambda_k$ is based on very few observations, and thus, very unstable. We come back to this issue in Section \ref {s:stab} of the appendix.




{\bf Remarks:} 1. It will be interesting to study the effect of noise according to Fourier degrees, and compute the effective readout error for noisy simulations of random quantum circuits (and other type of circuits) for both smaller and larger numbers of qubits and for various other noise models. In particular, this matter could be further checked for other values of $n$ and other types of circuits of the Google experiment. Of the three types of circuits from the Google experiments: full circuits, elided circuits, and patch circuits, we only considered here full circuits. It would be interesting to study the other types of circuits, and, in particular, the samples for patch circuits that were provided by the Google team in June 2022. Since patch circuits operate separately on two non-interacting sets of qubits, we may learn about the behavior of experimental data for smaller values of $n$, $6 \le n \le 12$.

2. In Section \ref {s:dolev} of the appendix we describe the Fourier behavior of data from a recent experiment \cite {Blu+23} of quantum computing based on neutral atom quantum device and rudimentary form of fault tolerance.




\newpage

\section {Fidelity and readout estimations: simulations and experimental data}

\label {s:fidelity}
In this section we use Fourier analysis to compute the best fit for the parameters $s$ and $q$ for our primary noise model given by Equation \eqref {e:noise-model-rho} both for the Google experimental data and for noisy simulations of circuits with $n$ qubits,  $n=12,14$. In a similar direction we compute our estimators for $\phi_{ro}$ 
for the Google data for circuits of size $n \le 36$. (In \cite {RSK22} we computed the estimator for $\phi_{ro}$ only for circuits with $n=12,14$.) We also compute the estimator for $\phi_{ro}$ 
for simulations of the experimental circuits with $n$ qubits, $n=12,14$. For these computations using the (fast)
Fourier--Walsh transform gave considerable computational advantage. 

To complement our analysis from \cite {RSK22} we  compute for simulations the estimators $U$, $V$, $MLE$ and $T$ as well as $S$, a modified version of $T$ (see \eqref{eq:S_squared}). (This part does not rely on Fourier analysis.) 

\subsection {Effective readout errors for the data and for simulations}
\label{s:sq}

We computed the parameters of the model described by \eqref {e:noise-model-rho}
$$sT_{(1-2q)}({\cal P}_C)+(1-s)(1/M),$$ that fit best the Google experimental data as well as various simulations. 

Recall that this is the symmetric readout noise model under the assumption that the gate errors are uncorrelated with readout errors. Under this assumption, $s=\phi_g$, the probability of the event ``no gate errors", and $q$ is the average readout error. We can expect that gate errors will push the estimated $q$ up beyond the estimate that is based on the noise model given by \eqref {e:symm-readout}, namely beyond the estimate based on the assumption that all gate errors take you to the uniform probability. Similarly, gate errors will push $s$ up beyond the value of $\phi_g$. We  refer to the value of $q$ that best fits the data as the ``effective readout error;" it accounts for the physical readout error plus some contribution of the gate errors which mathematically affect the samples like additional readout errors. 

Indeed, for the Weber QVM noise model,  for $n=12$, we have that on average $s=0.627$ and $q=0.053$, and for $n=14$ we have $s=0.532$ and $q=0.047$. It appears that the gate errors substantially increased the estimated value of $q$ in the Google simulations. The effective readout error is 39\% and 24\% higher respectively than the physical readout error which is 0.038.

For the Google experimental data we have that for $n=12$ on average $s=0.565$ and $q=0.035$, and for $n=14$ we have $s=0.510$ and $q=0.031$. For the values we get from the Google experimental data the effective readout error $q$ is indeed close (in fact, somewhat lower, by 8\% and 19\%, respectively) to the physical readout errors and there is no additional effect of gate errors.

We  also note that the coefficient of variation (standard deviation divided by the average) for $q$ and for $s$, is smaller for the Google experimental data than for the Weber QVM simulation. This is also the case for the MLE estimator for $\phi_{ro}$.\footnote {The coefficient of variation (CV),  is a  statistical measure of dispersion of a probability distribution.}

For more details see Section \ref{s:sq-simulations} where we also included  outcomes from the IBM Fake Guadalupe simulator, for a random circuit with 12 qubits, and from our 5-qubit random circuits runs on the Jakarta and Nairobi IBM freely-available (at the time) 7-qubit quantum computers.\footnote {The IBM quantum computers Guadalupe, Nairobi and Jakarta (see \cite {Wik}) are no longer available in IBM's fleet of quantum computers.} In Section \ref {s:dolev} we analyzed samples for a pair of 12-qubit logical circuits from a recent experiment \cite {Blu+23} of quantum computing based on neutral atoms. 

{\bf Remarks:} 1. In \cite {RSK22} we used MLE to find the best fit for the more detailed asymmetric noise model where  $q_1$ is the probability that 1 is read as 0  and $q_2$ is the probability that 0 is read as 1.
Quoting from \cite {RSK22}
``we computed the MLE
with $(\phi_g,q_1,q_2)$ as parameters, and obtained the estimates
(0.5571, 0.0465, 0.0196) for Google’s file 1." The actual readout values for the Google device are $q_1=0.055$ and $q_2=0.023$ and the estimation based on the samples are somewhat lower. 
Fast Fourier transform allowed us more efficient implementation of MLE in the symmetric case. We did not implement a similar improvement in the asymmetric case (or for the case that readout errors change for different qubits). 

2. We also performed the following simulation with Cirq. We used the Google experimental files with 12 and 14 qubits with 1-gate depolarization errors on all 1-gates (including those for the calibration), and adjusted the error rate so that the overall fidelity is similar to the Google experimental fidelity. The best fit for the parameters $s$ and $q$ in the model given by Equation \eqref {e:noise-model-rho} was (averaged on 10 simulation files) 
$s=0.4207$ and $q=0.013$, and in this case, the entire effective readout error comes from the depolarizing gate errors. See Figure \ref {fig:gates1}. 

3. The MLE estimation for $(s,q)$ can be seen as fitting equation \eqref {e:noise-model-rho} to the values $\Lambda_k$, $1 \le k \le n$. See, e.g., Figures \ref {fig:Fourier_Delta_Google_data}, \ref {fig:ohad12+14} and other figures throughout the paper. In this fitting the weight of $\Lambda_k$ is ${{n} \choose {k}}$.

\subsection{$\phi_{ro}$ estimators for the Google experimental data and for samples from Google's simulators}





\begin{figure}[h]
\centering
\includegraphics[width=\textwidth]{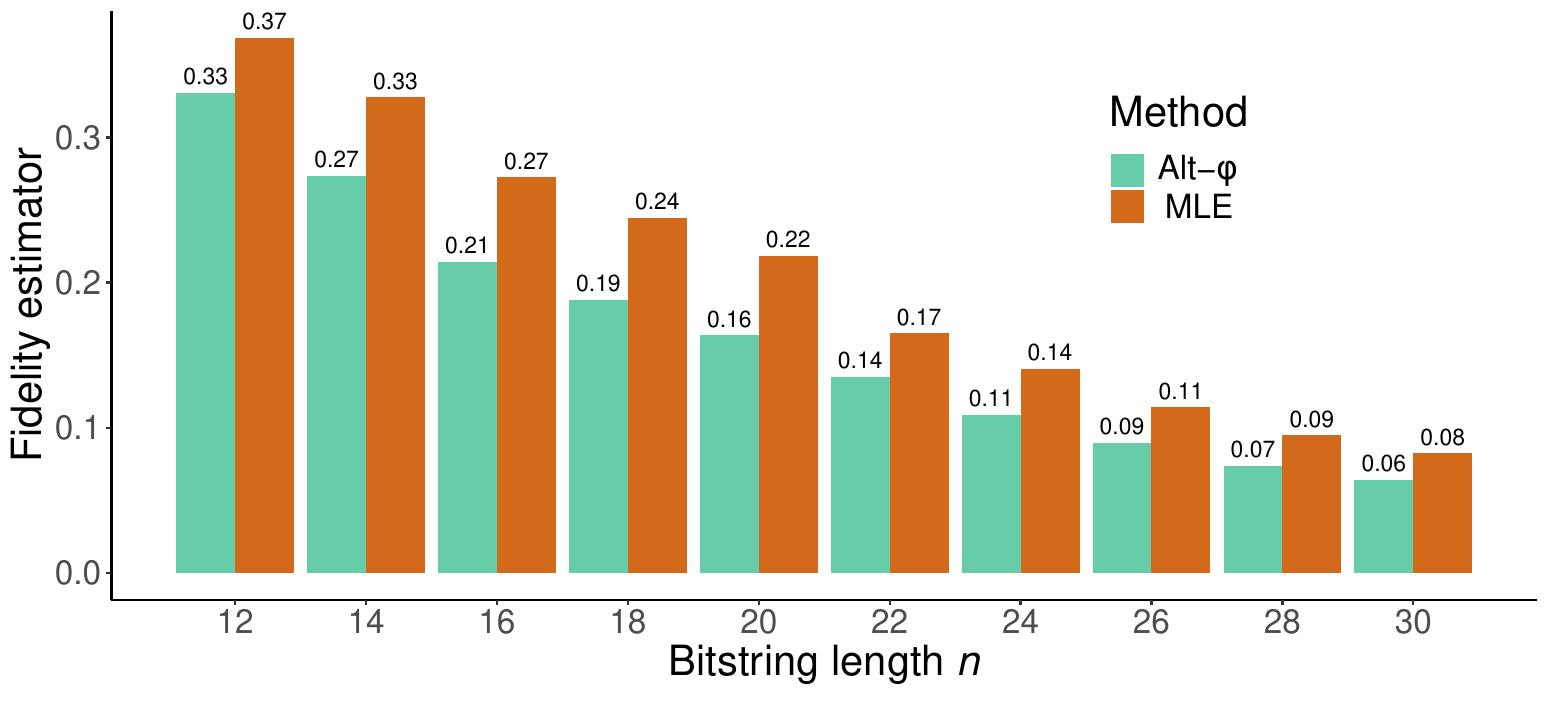}
\caption{The primary MLE fidelity estimator for the fidelity compared to the fidelity estimator alt-$\phi$ based on tracing the ``secondary signal" (Equation \eqref{e:secondary}) in the data. For alt-$\phi$ we use the estimator for $\phi_{ro}$ from Section 6 of \cite {RSK22} (also with the MLE statistics), and Formula \eqref{eq:phi_ro}. See Table \ref{t:dataFig6} for the data.}
\label{fig:fid-readout}
\end{figure}

In Section 6 of \cite {RSK22} we developed estimators for the probability $\phi_{ro}$ that there are no gate errors
and at least one readout error. This also allows an alternative estimator for the fidelity 
$\phi$ based on a ``secondary" signal. 

We repeat some basic notations. $\phi_g$ denotes the probability of no gate errors and $\phi_{ro}=\phi_g-\phi$ denotes the probability of the event that there are no gate errors and at least one readout error. Taking Formula (77) for granted,  gives us that 


   $$ \phi = \phi_g(1-q)^n,$$ 
This implies that 
\begin{equation}\label{eq:phi_ro}
 \phi_{ro} = \phi ((1-q)^{-n} - 1),   
\end{equation}

where for the Google experiment \cite {Aru+19} $q=0.038$. Our estimators for $\phi_{ro}$ are ``orthogonal" to the estimators for $\phi$ since they are based on the correlation between  the empirical distribution  the ``secondary signal" ${\cal N}_C^{ro}(x)$ (equation \eqref {e:secondary}). Estimating $\phi_{ro}$ allows an independent estimator for $\phi$ by Equation \eqref{eq:phi_ro}.  
As discussed in Section \ref {s:fourier-readout} above, these estimators could be computed using the Fourier--Walsh transform of the empirical data and the ideal probabilities. These estimators were based on the assumption that gate errors are uncorrelated to readout errors (the noise model \eqref {e:symm-noise-gates}).

For the experimental data from Google's quantum supremacy experiment, our estimates from \cite {RSK22} for $\phi_{ro}$ which are  based on the ``secondary" signal of the readout errors give values which are roughly 20\% lower than the values based on the primary estimated fidelity. See Figure \ref {fig:fid-readout} and Table  \ref{t:dataFig6}. In particular we do not witness an effect of gate errors toward higher estimated values of $\phi_{ro}$. 

For the Google's Weber QVM simulations, our estimates from \cite {RSK22} for $\phi_{ro}$ and hence also for alt-$\phi$, are  higher than the values based on the primary estimated fidelity. The estimated value for $\phi_{ro}$ (and hence also of alt-$\phi$) is larger (on average) by 38\% for $n=12$ and by 28\% for $n=14$ than the values based on the ``primary" signal. This appears to reflect the effect of gate errors. For more tables and discussion see Section \ref{s:sq-simulations}.




\subsection {Fidelity estimators for simulations}
\label {s:estimators}

We compared fidelity estimators that were studied in the Google paper \cite {Aru+19} and our paper \cite {RSK22} for the Google data and for simulations.
 Table \ref {t:QVM} summarizes the outcome of various estimators for the fidelity  for 
the ten experimental circuits for $n=12$ and $n=14$ for simulations with Google's ``Weber QVM" simulators. Table \ref {t:QVM} can be compared with Table \ref {t:google-experiment} in Section \ref {s:app.data.g} of the Appendix, that summarizes the outcome of various estimators for the fidelity  for the ten experimental circuits for $n=12$ and $n=14$ for the Google experimental data.




For the Weber QVM simulators the fidelity estimators $U$, $V$ and $MLE$ perfectly agree on average with the (77) predictors.\footnote {Carsten Voelkmann (private communication) asked if this agreement weakens the ``too good to be true” criticism of Formula (77) from \cite {KRS23,Kal23}. We note that the agreement for simulation has no bearing on the main reasons for concern from Section 4.3 of \cite {KRS23}, but that we do not witness here the systematic difference between ${\cal F}_{XEB}$ and the probability for the event of no errors which is anticipated in \cite {Gao+21} for large values of $n$.} 


Table \ref {t:fake-guadalupe} in the appendix summarizes the outcome of various estimators for the fidelity $\phi$ for ten 12-qubit random circuits for 
simulations with IBM's ``Fake Guadalupe" simulator. We note that the ideal distributions 
for the IBM circuits are pretty far from ideal Porter–Thomas distributions, and consequently MLE
(and $V$) give considerably better estimations compared to $U$. In this case, there is a
substantial gap (28\% and 40\%) between the averaged estimations by $V$ and $MLE$ and the prediction from Formula (77).

\begin{table}
\begin{center}\resizebox{1.0\textwidth}{!}{
 \begin{tabular}{||c c c c c c c c c c||}
 \hline
 $n$&File&(77)&$U$&$V$&MLE&$T$&$S$&alt-$\phi$&$\phi_{ro}$\\
 \hline\hline 
12&0&0.324&0.338&0.332&0.334&0.427&0.424&0.611&0.362\\
12&1&0.324&0.327&0.322&0.321&0.415&0.412&0.406&0.240\\
12&2&0.324&0.318&0.324&0.325&0.411&0.415&0.398&0.236\\
12&3&0.324&0.325&0.322&0.319&0.415&0.413&0.479&0.284\\
12&4&0.324&0.336&0.323&0.323&0.423&0.415&0.460&0.273\\
12&5&0.324&0.344&0.333&0.332&0.430&0.423&0.575&0.340\\
12&6&0.324&0.307&0.314&0.317&0.403&0.407&0.275&0.163\\
12&7&0.324&0.346&0.320&0.321&0.423&0.407&0.417&0.247\\
12&8&0.324&0.314&0.332&0.328&0.415&0.427&0.475&0.281\\
12&9&0.324&0.312&0.319&0.324&0.407&0.412&0.368&0.218\\
12&\textbf{Avg}&&\textbf{0.327}&\textbf{0.324}&\textbf{0.324}&\textbf{0.417}&\textbf{0.416}&\textbf{0.446
}&\textbf{0.264}\\
12&\textbf{Std}&&\textbf{0.013}&\textbf{0.006}&\textbf{0.005}&\textbf{0.008}&\textbf{0.007}&\textbf{0.093}&\textbf{0.055}\\
\hline
\multicolumn{8}{c}{} \\ \hline
\hline
 $n$&File&(77)&$U$&$V$&MLE&$T$&$S$&alt-$\phi$&$\phi_{ro}$\\
 \hline\hline 
14&0&0.269&0.279&0.271&0.269&0.382&0.376&0.337&0.243\\
14&1&0.269&0.270&0.269&0.270&0.381&0.380&0.349&0.251\\
14&2&0.269&0.266&0.267&0.266&0.373&0.374&0.289&0.208\\
14&3&0.269&0.271&0.269&0.269&0.380&0.379&0.334&0.241\\
14&4&0.269&0.268&0.273&0.273&0.380&0.383&0.435&0.313\\
14&5&0.269&0.270&0.268&0.266&0.379&0.378&0.333&0.240\\
14&6&0.269&0.273&0.266&0.266&0.377&0.373&0.361&0.260\\
14&7&0.269&0.262&0.261&0.262&0.371&0.370&0.234&0.169\\
14&8&0.269&0.276&0.275&0.275&0.387&0.386&0.426&0.307\\
14&9&0.269&0.265&0.266&0.266&0.378&0.379&0.331&0.238\\
14&\textbf{Avg}&&\textbf{0.270}&\textbf{0.269}&\textbf{0.268}&\textbf{0.379}&\textbf{0.379}&\textbf{0.343}&\textbf{0.247}\\
14&\textbf{Std}&&\textbf{0.005}&\textbf{0.004}&\textbf{0.004}&\textbf{0.004}&\textbf{0.005}&\textbf{0.056}&\textbf{0.040}\\

    \hline
\end{tabular}}
\caption {The various estimators for $\phi$ for the 10 files with $n=12,14$ for Weber QVM simulations. Here, the simulation itself and the (77) values are based, for individual components, on average fidelity values from the Google 2019 experiment. (For readout errors we use the “symmetric” average readout error (0.038).) The fidelity estimators $U$, $V$ and $MLE$ perfectly agree on average with the (77) predictors. For individual circuits, in agreement with \cite {RSK22}, $U$ has larger deviations. 
The value of alt-$\phi$ is larger (on average) by 38\% (for $n=12$) and 28\% (for $n=14$) than the MLE values and this may reflect the effect of gate errors on the decrease of Fourier--Walsh coefficients. Here we use the ${\cal F}_{XEB}$-style estimator for $\phi_{ro}$; using the MLE estimation gives similar averaged values and smaller deviations (Table \ref {t:mle-phiro-sim}).}
\label {t:QVM}
\end{center}

\end {table}







\newpage

\section {Concluding remarks}
\label{s:conc}
In this paper we developed estimators $\Lambda_k,$ $ k=1,2,\dots,n$ for the contribution of degree $k$ Fourier--Walsh coefficients for empirical data coming from random circuit sampling.
We studied a two-parameter noise model given by 
$$ 
s T_{(1-2q)} ({\cal P}_C(x)) + \frac{1-s}{2^n}~~~~~0\le s\le 1; ~0 \le q\le 1/2,
$$
and used Fourier methods for efficient estimation of its parameters and for computing 
other estimators from \cite {RSK22}. Using these tools we studied the behavior of samples from the Google quantum supremacy experiment \cite {Aru+19} and compared the behavior of the Google data with Google's (and other) simulators.  Our estimators $\Lambda_k$, the general two-parameter noise model, 
and other ingredients of our study could be relevant to the study of data from various other NISQ experiments and simulations, and for analysis and modeling of noise. 

Generally speaking, the effect of noise is to suppress exponentially the high degree  Fourier--Walsh coefficients. For the data we studied, the dominant effect is that of readout errors and this is exhibited both for the experimental data and for simulations. According to simulations on the Google simulator, a smaller additional effect of suppressing high-degree Fourier coefficients comes from gate errors.  This effect of gate errors is also expected from theoretical studies such as Aharonov et al. \cite {AGLLV22}, and it is not observed in the Google experimental data.

\subsubsection *{The Google data compared to simulations}

For the data coming from Google's supremacy experiment, the decay of the Fourier contribution according to Fourier degrees is  similar to what could be expected from the effect of readout errors with no additional effect of gate errors. In other words, the samples from the Google experiment are close in this respect to simulations with our modeling of readout errors (Equation \eqref {e:symm-readout}) where
every gate error moves us to the uniform distribution on bitstrings. This is different from the decay of Fourier coefficients for the Google simulations: there the gate errors have additional contribution to the decay of higher degree Fourier coefficients.

A similar quantitative picture is shown by our estimators for readout errors from Section 6 of \cite {RSK22}. 
Our estimator for $\phi_{ro}$ based on the secondary signal given by Formula \eqref {e:secondary} 
gives (for $12 \le n \le 30$) values which are roughly 20\% lower than the primary estimated fidelity. 
In contrast, for Google's noisy simulators and $n=12,14$ gate errors increase the estimated value of 
$\phi_{ro}$ by 38\% and 28\%, respectively.  Comparing the effective readout errors (namely the estimated value of $q$) between simulations and the experimental data is similar.  
For $n=12,14$ the effective readout error for the experimental data is somewhat lower (0.035 and 0.031, respectively) compared to the physical readout error (0.038). For the simulations, the effective readout error is larger (0.053 and 0.047, respectively) than the prescribed readout error. 

We also note that for our estimators for $s$ and $q$, and for the MLE estimator for 
$\phi_{ro}$, 
the coefficient of variation (standard deviation divided
by the average) for the different circuits is 
smaller for Google's experimental data compared to the data coming from Google's simulator. 

We studied the effect of gate errors on the Fourier behavior, and on estimations of the $(s,q$)-pairs also for samples coming from the IBM ``Fake Guadalupe" simulator. The additional effect of gate errors toward larger values of $s$ and $q$ was stronger than that of samples from the Google simulator. 





\subsubsection *{Simulations of noisy quantum circuits}

The work of this paper was our first experience with simulations of noisy quantum circuits and we used both the Google and IBM simulators. Our main objective in this paper was to use simulations for studying the effect of gate errors on the Fourier coefficients, but simulations of noisy random circuits gave an opportunity to examine other matters, and overall simulations for noisy quantum circuits confirm some of the lessons from our statistical studies of the fidelity estimators in \cite {RSK22}.  In particular, the MLE and $V$ fidelity estimators are consistently advantageous compared to ${\cal F}_{XEB}$. This is especially the case when the noiseless simulation is further away from a Porter--Thomas distribution, such as for IBM's random circuits. 

\subsubsection *{Comparing IBM and Google}

We also studied data from other quantum computer experiments and even ran ourselves a few five-qubit random circuits on 7-qubit IBM quantum computers. We are aware of some random circuit sampling experiments on the IBM quantum computers with six qubits \cite {Kim+21} and seven qubits (S. Merkel, private communication, 2023).\footnote {We could not obtain the samples for these 6-qubit and 7-qubit experiments.}

Note that the  difference between Google's empirical assertions for random circuits with 12--53 qubits and the current situation with random circuit sampling on IBM quantum computers is quite substantial. While IBM  offered  online quantum computers in the range between 7 and 133 qubits, we are not aware of random circuit experiments on those IBM computers with more than seven qubits.\footnote{We do not know if running on the actual IBM Guadalupe quantum computer (or some other IBM quantum computer) 12-qubit circuits  similar to those we simulated on the Fake Guadalupe simulator can lead to samples of fair quality or even to quality similar to the simulations.} 

 


\subsubsection* {The boson sampling experiment}
In 2020, a quantum
advantage claim based on boson sampling was presented \cite {Zho+20} by a group from the
University of Science and Technology of China, using a different
technology and generating a different distribution,  but sharing some statistical
principles with Google's demonstration. In this case, the probabilities are described
by a function $f$ of Gaussian complex variables and a variant of the Fourier expansion called
Fourier--Hermite expansion can be used to analyze boson sampling distributions \cite{KalKin14}. 
(See further discussion and references in \cite {KalKin21}.)

It could be possible to employ a statistical approach similar to the one presented in this paper to estimate the impact of noise on Fourier--Hermite coefficients for boson sampling experiments.

\subsubsection* {Improved classical algorithms via Fourier expansion}

As first pointed out in Kalai and Kindler \cite {KalKin14} for boson sampling experiments,
low degree Fourier--Hermite approximation is expected to give good classical approximation algorithms when the fidelity is of order $1/n$ and $n$ is the number of bosons. The reason is
that a substantial $\ell_2$-part of the distribution is supported on low Fourier--Hermite degrees. Kalai--Kindler's work \cite {KalKin14} (and later works) largely refute the huge advantage claim of the boson sampling experiment from \cite {Zho+20} (and from similar works). See also \cite {OJF23} for a recent Fourier-based classical algorithm for noisy boson sampling.

Boixo, Smelyanskiy, and Neven \cite {BSN17} pointed out that the situation is different for random circuit sampling since for Porter--Thomas distributions we can expect that all Fourier coefficients represent identical Gaussian behavior.
Still, the Fourier description supports the heuristics (proposed in the Google paper \cite {Aru+19}) that for random circuit sampling, the computational complexity of classical approximation
algorithms will behave proportionally to the fidelity. Aharonov et al. \cite 
{AGLLV22} used
an algorithm based on low degree Fourier coefficients to give a 
polynomial time algorithm for approximating samples obtained by noisy random circuit sampling experiments under a certain noise model.

\subsubsection* {
Kalai and Kindler's 
interpretation of the Fourier behavior}

Kalai and Kindler's interpretation of their results from \cite {KalKin14} was: ``These results seem to weaken the possibility of demonstrating quantum-speedup via boson sampling without quantum fault-tolerance." Kalai \cite {Kal18,Kal20,Kal23} extended this interpretation to all NISQ computers and argued that impossibility of NISQ computers to demonstrate quantum supremacy would weaken the possibility of NISQ computers to achieve good-quality quantum error correction (a task that appears harder than quantum supremacy).

The interpretation of the Fourier behavior as an obstacle for capabilities of NISQ systems is based on two ingredients.

1) {\bf Noise sensitivity:} for a large range of error-rates the correlation between the empirical distribution to the ideal distribution tends to zero. Moreover, the empirical distribution will be very sensitive  to small changes in the mathematical description of the noise itself, leading to inherently non-stationary and inherently unpredictable empirical distributions.

2) {\bf Computational complexity:} for a fixed error-rate, noisy distributions of NISQ systems represent low level computational complexity class.

Both these ingredients could be relevant to statistical analysis of data coming from NISQ computers.

\subsubsection *{Relevant recent experiments}

It would be interesting to apply statistical tools of this paper and our earlier ones \cite {RSK22,KRS22d,KRS23} to other experiments for NISQ computers. In addition to the boson sampling experiments \cite {Zho+20} and the IBM 6-qubit RCS experiment \cite {Kim+21} mentioned above, there are additional experiments  for which our tools could be relevant, Section \ref{s:nisqexp} 
in the appendix lists a few notable recent NISQ experiments, among those we study (very partial) data coming from a recent study of logical circuits based on neutral atoms.  

\subsubsection *{A recent experiment with rudimentary form of quantum fault-tolerance} 

A recent 2024 experiment \cite {Blu+23} by a group from Harvard, MIT, and QuEra, drew considerable attention. It describes logical circuits based on neutral atoms. We used our statistical tools to analyze samples of  12 qubit logical circuits from \cite {Blu+23}. (In this particular experiment, eight physical experiment encoded three logical qubits. This is one out of many experiments described in the paper.) We were pleased to see that our programs applied smoothly and gave relevant analysis. The study 
of fidelity estimators and Fourier contributions is presented in Section \ref {s:dolev} of the appendix and we note the stable nature of the Fourier behavior of the samples.

\subsection *{Acknowledgements}
Research supported by ERC grant 834735. We are thankful to our team-member Ohad Lev for doing a great job of running simulations on the IBM and Google simulators, running 5-qubit experiments on IBM quantum computers (Nairobi and Jakarta), and for fruitful discussions. We thank Seth Merkel for helpful discussion, and for providing the Qiskit program for random circuits (that we applied for our 5-qubit experiments, and that Merkel applied for 7-qubit circuits.) We thank Dolev Bluvstein for helpful discussion and for providing data for 12 logical qubit IQP state described in \cite {Blu+23}. We also thank Carsten Voelkmann for many helpful corrections and thoughtful suggestions.


\newpage

\section {Appendix A: additional data}
\label {s:ad}
In this section we present additional data based on Google's 2019 experiment and based on our simulations with the Google simulator Weber QVM, data coming from simulations with IBM's simulator ``Fake Guadalupe", and data from our 5-qubit experiments on IBM 7-qubit quantum computers. 
\subsection{Estimating the couple $(s, q)$}
\label{s:sq-simulations}
A basic noise model for our study, going back to Section 6 of \cite {RSK22} is given by
$$s T_{(1-2q)}({\cal P}_C(x)) +(1-s) (1/M).$$ 
(See the discussion in Section \ref {s:sq}.) 
We used MLE to estimate the couple $(s,q)$ and we summarize the outcomes for
data coming from Google's quantum supremacy experiment, Google's simulator, IBM's simulator, and our 5-qubit random circuit experiments. 

\subsubsection *{Google experimental data}

\begin{table}[ht]
    \centering
\begin{tabular}[t]{||c c c c||} 
     \hline
    $n$&$i$&$s$&$q$\\[0.5ex] 
    \hline\hline 
    12&0&0.605&0.039\\
    12&1&0.555&0.033\\
    12&2&0.517&0.030\\
    12&3&0.570&0.034\\
    12&4&0.570&0.035\\
    12&5&0.648&0.046\\
    12&6&0.569&0.036\\
    12&7&0.518&0.030\\
    12&8&0.559&0.035\\
    12&9&0.541&0.031\\
    \textbf{12}&\textbf{Avg}&\textbf{0.565}&\textbf{0.035}\\
    \textbf{12}&\textbf{Std}&\textbf{0.037}&\textbf{0.004}\\
    \hline 
\end{tabular}\hfill%
\begin{tabular}[t]{||c c c c||} 
     \hline
    $n$&$i$&$s$&$q$\\[0.5ex] 
    \hline\hline 
    14&0&0.510&0.031\\
    14&1&0.561&0.038\\
    14&2&0.508&0.031\\
    14&3&0.506&0.030\\
    14&4&0.530&0.034\\
    14&5&0.462&0.025\\
    14&6&0.484&0.028\\
    14&7&0.490&0.028\\
    14&8&0.521&0.033\\
    14&9&0.533&0.034\\
    \textbf{14}&\textbf{Avg}&\textbf{0.510}&\textbf{0.031}\\
    \textbf{14}&\textbf{Std}&\textbf{0.027}&\textbf{0.004}\\
    \hline 
\end{tabular}
\caption{For Google's experimental data, $n=12,14$  gate errors do not increase the estimated values of $q$ and $s$. The averaged estimated value of $q$ is 0.035 for $n=12$ and $0.031$ for $n=14$, while the average readout error is $0.038$.}
\end{table}

\subsubsection *{Google's simulator Weber QVM} 

\begin{table}[ht]
\begin{tabular}[t]{||c c c c||} 
    \hline
    $n$&$i$&$s$&$q$\\[0.5ex] 
    \hline\hline 
    12&0&0.778&0.069\\
    12&1&0.629&0.054\\
    12&2&0.606&0.050\\
    12&3&0.525&0.041\\
    12&4&0.709&0.063\\
    12&5&0.782&0.070\\
    12&6&0.496&0.036\\
    12&7&0.586&0.049\\
    12&8&0.599&0.050\\
    12&9&0.560&0.045\\
    \textbf{12}&\textbf{Avg}&\textbf{0.627}&\textbf{0.053}\\
    \textbf{12}&\textbf{Std}&\textbf{0.094}&\textbf{0.011}\\
    \hline 
\end{tabular}\hfill%
\begin{tabular}[t]{||c c c c||} 
 \hline
$n$&$i$&$s$&$q$\\[0.5ex] 
\hline\hline 
14&0&0.512&0.045\\
14&1&0.542&0.049\\
14&2&0.471&0.040\\
14&3&0.606&0.056\\
14&4&0.576&0.053\\
14&5&0.518&0.046\\
14&6&0.533&0.049\\
14&7&0.419&0.033\\
14&8&0.630&0.058\\
14&9&0.511&0.046\\
\textbf{14}&\textbf{Avg}&\textbf{0.532}&\textbf{0.047}\\
\textbf{14}&\textbf{Std}&\textbf{0.059}&\textbf{0.007}\\
\hline 
\end{tabular}
\caption{For Google's Weber QVM simulation data, $n=12,14$  gate errors appear to  increase the estimated values of $q$ and $s$. The averaged estimated value of $q$ (that we referred to as the effective readout error) is 0.053 for $n=12$ and $0.047$ for $n=14$, while the physical readout error is 0.038.}
\end{table}

\subsubsection *{Comparison between Google's experiment and Google's simulation}

For Google's quantum supremacy experiment the averaged estimated value of $q$ (that we referred to as the effective readout error) is 0.035 for $n=12$ and 0.031 for $n=14$ and is close to the physical readout error (0.038). For Google's Weber QVM simulation the estimated value of $q$ is 0.053 for $n=12$ and 0.047 for $n=14$ and this appears to reflect a substantial effect of gate errors on the estimated value of $q$ which is not observed in the experiment itself. 
The estimated values of $s$ for the experiment are close to the value of $\phi_g$ as predicted by the fidelity and the rate of readout errors without additional contribution of gate errors. The estimated value of $s$ for the simulation appears to reflect also some contribution of gate errors.  

The coefficient of variation (CV) for the estimated values of $s$ and $q$ over the ten files is smaller for the experiment compared to the simulation: For $n=12$, for $s$ it is 0.065 for the experiment and 0.15 for the simulation, and for $q$ it is 0.114 for the experiment and 0.208 for the simulation. For $n=14$ the coefficient of variation for $(s,q)$  is (0.053, 0.130) for the experiment and (0.111, 0.145) for the simulation. One explanation for the larger CV value could be that in the simulations, the values of $q$ and $s$ are affected by gate errors and this effect could be different for different circuits leading to higher dispersion. In contrast, for the Google experimental data we do not witness the effect of gate errors on higher values of $q$ and $s$. We note, however, that there could be various reasons that may lead to a higher CV values for experimental data compared to simulations. 

\subsubsection*{Fake Guadalupe}

\begin{table}[ht]
\centering
 \begin{tabular}[t]{||c c c c||} 
 \hline
$n$&$i$&$s$&$q$\\[0.5ex] 
\hline\hline 
12&0&0.988&0.073\\
12&1&0.980&0.065\\
12&2&1.000&0.075\\
12&3&0.989&0.069\\
12&4&0.823&0.056\\
12&5&0.974&0.069\\
12&6&0.934&0.062\\
12&7&0.929&0.063\\
12&8&0.890&0.067\\
12&9&0.938&0.064\\
\textbf{12}&\textbf{Avg}&\textbf{0.945}&\textbf{0.066}\\
\textbf{12}&\textbf{Std}&\textbf{0.052}&\textbf{0.005}\\
\hline 
\end{tabular}
\caption {Estimating $s$ and $q$ for the IBM simulations. While the average (physical) readout errors is 0.02176, the effective readout error, namely the estimated value of $q$ from the data is 200\% higher. 
}
\label{table:fake_s_q}
\end{table}
For the Fake Guadalupe IBM simulator the averaged estimated value of $q$ is 0.066 which is three times larger than the average readout error. The average value of $s$ is rather close to 1.

\subsubsection*{Random circuits 5-qubit experiments on IBM 7-qubit Quantum computers}

\begin{table}[ht]
\centering
    \begin{tabular}[t]{||c c c c||} 
    \multicolumn{4}{c}{\textbf{Nairobi}} \\
    \hline
    $n$&$i$&$s$&$q$\\[0.5ex] 
    \hline\hline 
    5&0&1.000&0.083\\
    5&1&0.635&0.000\\
    5&2&0.732&0.024\\
    5&3&1.000&0.089\\
    5&4&0.945&0.053\\
    5&5&0.721&0.033\\
    5&6&1.000&0.060\\
    5&7&1.000&0.078\\
    5&8&0.768&0.018\\
    5&9&0.731&0.001\\
    \textbf{5}&\textbf{Avg}&\textbf{0.853}&\textbf{0.044}\\
    \textbf{5}&\textbf{Std}&\textbf{0.140}&\textbf{0.032}\\
    \hline 
    \end{tabular}\hfill%
    \begin{tabular}[t]{||c c c c||} 
    \multicolumn{4}{c}{\textbf{Jakarta}} \\
    \hline
    $n$&$i$&$s$&$q$\\[0.5ex] 
    \hline\hline 
    5&0&0.851&0.067\\
    5&1&1.000&0.095\\
    5&2&0.896&0.088\\
    5&3&0.793&0.063\\
    5&4&0.685&0.058\\
    5&5&0.946&0.109\\
    5&6&0.855&0.073\\
    5&7&0.550&0.021\\
    5&8&0.566&0.012\\
    5&9&0.867&0.083\\
    \textbf{5}&\textbf{Avg}&\textbf{0.801}&\textbf{0.067}\\
    \textbf{5}&\textbf{Std}&\textbf{0.145}&\textbf{0.029}\\
    \hline 
    \end{tabular}
\caption{Estimating the values of $s$ and $q$ for our 5-qubit experiment on the Nairobi and Jakarta IBM quantum computers.}
\end{table}

For our 5-qubit experiment on the IBM quantum computers Nairobi and Jakarta it appears that the estimated values of $s$ and $q$ are effected not only by readout errors but also by gate errors. It would be interesting to have similar experiments 
on larger circuits on IBM quantum computers of greater quality.  

\subsection {IBM simulations: the Fourier estimator $\Lambda_k$ }

\begin{figure}[h]
\centering
\includegraphics[width=\textwidth]{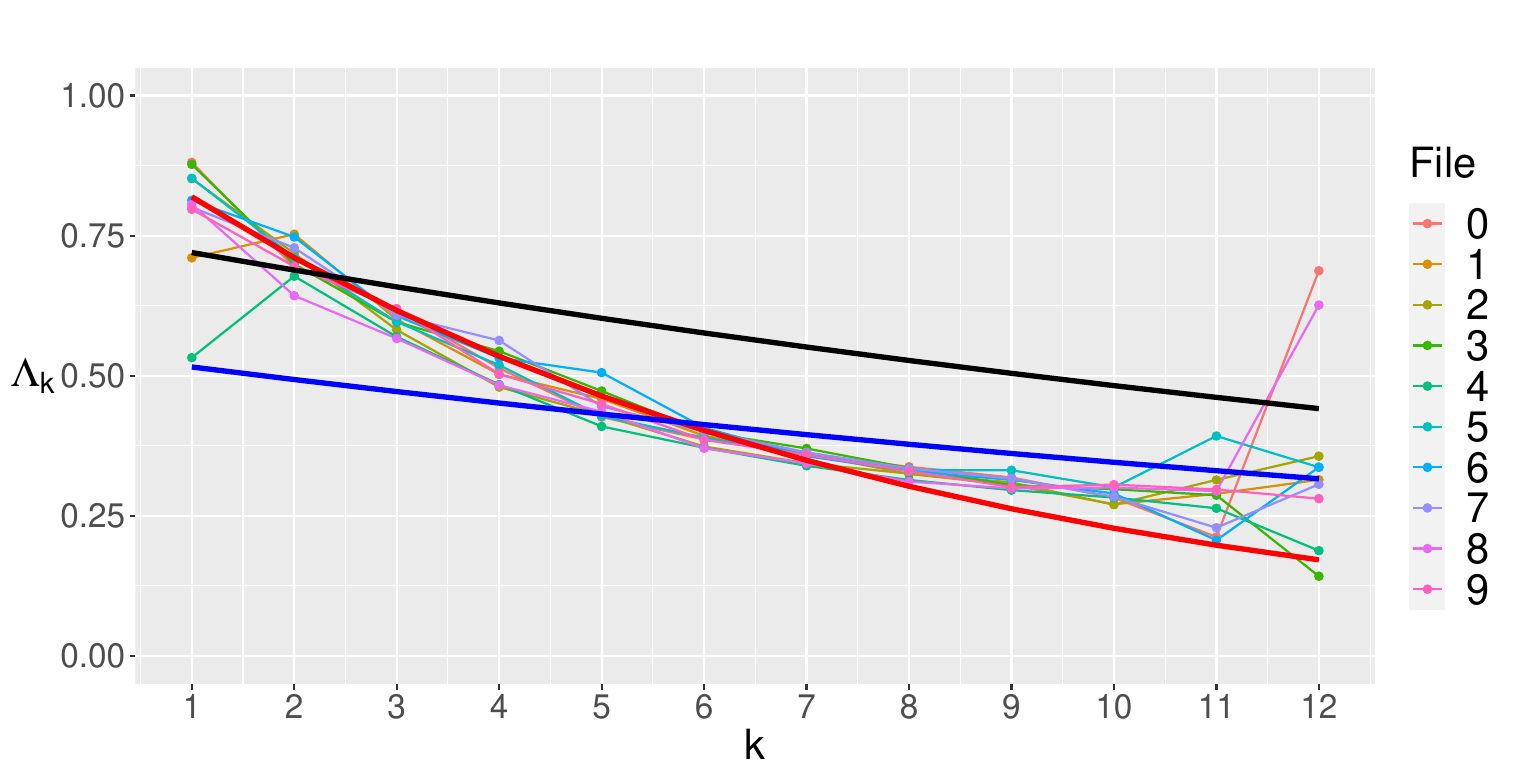}
\caption{The decay of the relative contribution of degree-$k$ Fourier coefficient degree $k$ Fourier coefficients for 12-qubit random circuits on IBM's ``Fake Guadalupe" simulators with IBM's full noise model for their 16-qubit quantum computer Guadalupe.   The solid black curve describes the effect of readout errors \eqref{eq:reference_readout} assuming that every gate error leads to a uniform random bitstring. It is based on $q=0.022$, $\phi_g=0.753$ and these values were derived from the average MLE, see Table \ref{t:fake-guadalupe}, using Equation \refeq{e:phiro-ibm}. The solid blue curve is based on $q=0.022$, and $\phi_g=0.541$ that is derived from the prediction by Formula (77). 
The solid red curve is based on the best fit $s=0.945$ and $q=0.066$, given in Table \ref{table:fake_s_q}.}
\label{fig:Fourier-fake-guadalupe}
\end{figure}

The decay of the relative contribution of degree-$k$ Fourier coefficient degree $k$ Fourier coefficients for 12-qubit random circuits on IBM's ``Fake Guadalupe" simulators with IBM's full noise model for their 16-qubit quantum computer Guadalupe is shown in Figure \ref{fig:Fourier-fake-guadalupe}. While the physical readout error is 0.022, the effective readout error of $q=0.066$ is considerably larger. 

\subsection {IBM data: fidelity estimators}
In Table \ref {t:fake-guadalupe} we describe various fidelity estimators for the data gathered from running ten random circuits with 12 qubits on the Fake Guadalupe that provides a simulation for IBM's 16-qubit Guadalupe quantum computer. We note that the distributions of probabilities are not close to Porter--Thomas distributions and this can explain the large fluctuation in the value of $U$ (=${\cal F}_{XEB}$). In fact, its average is greater than 1. As expected, MLE and $V$ appear to be more stable and closer to the value given by the prediction of Formula (77). The normalized version of $T$, referred to as $S$ gives close values to MLE.

\begin{table}
\begin{center}\resizebox{\textwidth}{!}{
 \begin{tabular}{||c c c c c c c c c c||} 
 \hline
 $n$&File&(77)&$U$&$V$&MLE&$T$&$S$&alt-$\phi$&$\phi_{ro}$\\ [0.5ex] 
 \hline\hline 
12&0&0.414&0.960&0.554&0.568&0.787&0.598&1.301&0.398\\
12&1&0.414&1.701&0.543&0.623&1.020&0.576&1.428&0.437\\
12&2&0.414&0.894&0.511&0.548&0.735&0.556&1.281&0.392\\
12&3&0.414&1.945&0.567&0.657&1.114&0.601&1.370&0.419\\
12&4&0.414&1.466&0.493&0.547&0.897&0.520&1.346&0.412\\
12&5&0.414&1.331&0.525&0.565&0.895&0.562&1.350&0.413\\
12&6&0.414&1.305&0.549&0.575&0.902&0.585&1.415&0.433\\
12&7&0.414&1.221&0.551&0.581&0.872&0.585&1.389&0.425\\
12&8&0.414&0.871&0.491&0.512&0.700&0.525&1.265&0.387\\
12&9&0.414&1.721&0.542&0.603&1.026&0.575&1.386&0.424\\
12&\textbf{Avg}&&\textbf{1.342}&\textbf{0.533}&\textbf{0.578}&\textbf{0.895}&\textbf{0.568}&\textbf{1.352}&\textbf{0.414}\\
12&\textbf{Std}&&\textbf{0.352}&\textbf{0.025}&\textbf{0.039}&\textbf{0.125}&\textbf{0.026}&\textbf{0.054}&\textbf{0.016}\\
     \hline 
\end{tabular}}
\caption {The various estimators for $\phi$ for the the 10 files with $n=12$ for IBM's Fake Guadalupe simulations. We note that the ideal distributions are pretty far from ideal Porter--Thomas distributions, and consequently MLE (and $V$) give considerably better estimations compared to $U$. 
There is a substantial gap between $V$ and MLE and the prediction from Formula (77). The value of $V$ is 28\% higher, and the MLE value is 40\% higher. 
Unlike the Google experimental data and simulations, $S$ is quite close to MLE. 
The estimates for $\phi_{ro}$ is substantially larger than $0.306 \phi$ 
(Formula \eqref {e:phiro-ibm}) which is based on average readout error 0.22. (Subsequently, the values of {\it alt}-$\phi$ are larger than 1.) This may account for the effect of gate errors. }  
\label {t:fake-guadalupe}
\end {center}
\end {table}

For the IBM data we witnessed a large gap between the (physical) averaged readout error and the effective readout error. For the Fake Guadalupe simulation the average readout error is 0.022 so  we expect that $\phi= \phi_g(1-0.022)^{12}$. From $\phi_g=\phi+\phi_{ro}$ 
we obtain that \begin {equation}
\label{e:phiro-ibm}
\phi_{ro} \approx 0.306 \phi.
\end {equation}
(The alternative estimate for the fidelity itself based on the secondary signal is {\it alt}-$\phi=\phi_{ro}/0.303$.) The actual estimated values of $\phi_{ro}$ are considerably larger 
than $0.306 \phi$ and this may account to the effect of gate errors. 




Tables \ref{t:nairobi} and \ref {t:jakarta} provide fidelity estimates for our experiments with 5-qubit random circuits on IBM's quantum computers Nairobi and Jakarta, respectively. Also here the value of ${\cal F}_{XEB}$ is more volatile than MLE and $V$. For the Nairobi experiment the (77) prediction for the fidelity is 0.719 and the prediction for $\phi_g$, the event of ``no gate errors" is 0.819, and therefore $\phi_{ro}=0.100$.  It follows that $\phi_{ro}=0.140 \phi$. Our estimators for $\phi_{ro}$ give much higher values. For the Jakarta experiment the (77) prediction for the fidelity is 
0.736, and for $\phi_g$ is 0.834.  

We note that in 
Merkel's 7-qubit random circuit experiment (private communication, 2023) we also witnessed volatile behavior of ${\cal F}_{XEB}$ with values sometimes greater than one. (In this case we do not have the data for computing $V$ and $MLE$.)  

\begin {table}

\begin{center}\resizebox{\textwidth}{!}{
 \begin{tabular}{||c c c c c c c c c c||} 
 \hline
 $n$&File&(77)&$U$&$V$&MLE&$T$&$S$&alt-$\phi$&$\phi_{ro}$\\ [0.5ex]
 \hline\hline 
5&1&0.719&0.801&0.595&0.607&0.777&0.649&&0.398\\
5&2&0.719&0.401&0.659&0.635&0.654&0.813&&0.437\\
5&3&0.719&0.626&0.598&0.642&0.684&0.647&&0.392\\
5&4&0.719&0.504&0.583&0.613&0.618&0.644&&0.419\\
5&5&0.719&0.794&0.852&0.754&0.912&0.915&&0.412\\
5&6&0.719&0.373&0.576&0.631&0.575&0.693&&0.413\\
5&7&0.719&0.695&0.768&0.757&0.809&0.824&&0.433\\
5&8&0.719&0.294&0.672&0.626&0.627&0.920&&0.425\\
5&9&0.719&0.713&0.644&0.687&0.746&0.687&&0.387\\
5&10&0.719&0.714&0.696&0.729&0.863&0.826&&0.424\\
&\textbf{Avg}&&\textbf{0.592}&\textbf{0.664}&\textbf{0.668}&\textbf{0.726}&\textbf{0.762}&\textbf{$\bf \gg 1$}&\textbf{0.414}\\
&\textbf{Std}&&\textbf{0.180}&\textbf{0.088}&\textbf{0.055}&\textbf{0.102}&\textbf{0.108}&\textbf{}&\textbf{0.017}\\
     \hline 
\end{tabular}}
\end{center}
\caption {Fidelity estimators for our 5-qubit experiments with IBM's 7-qubit Nairobi quantum computer.
} 
\label {t:nairobi}
\end {table}

\begin {table}
\begin{center}\resizebox{\textwidth}{!}{
 \begin{tabular}{||c c c c c c c c c c||} 
 \hline
 $n$&File&(77)&$U$&$V$&MLE&$T$&$S$&alt-$\phi$&$\phi_{ro}$\\ [0.5ex] 
 \hline\hline 
5&1&0.736&0.608&0.601&0.647&0.715&0.689&&0.370\\
5&2&0.736&0.256&0.586&0.603&0.452&0.662&&0.302\\
5&3&0.736&0.555&0.575&0.601&0.639&0.630&&0.297\\
5&4&0.736&0.446&0.555&0.572&0.553&0.598&&0.363\\
5&5&0.736&0.353&0.519&0.501&0.492&0.578&&0.334\\
5&6&0.736&0.773&0.513&0.559&0.715&0.565&&0.237\\
5&7&0.736&0.514&0.539&0.567&0.601&0.597&&0.344\\
5&8&0.736&0.479&0.456&0.499&0.591&0.559&&0.426\\
5&9&0.736&0.409&0.519&0.525&0.517&0.564&&0.490\\
5&10&0.736&0.949&0.525&0.579&0.781&0.563&&0.307\\
&\textbf{Avg}&&\textbf{0.534}&\textbf{0.539}&\textbf{0.565}&\textbf{0.605}&\textbf{0.600}&\textbf{$\bf \gg 1$}&\textbf{0.068}\\
&\textbf{Std}&&\textbf{0.142}&\textbf{0.042}&\textbf{0.047}&\textbf{0.088}&\textbf{0.044}&\textbf{}&\textbf{0.070}\\
     \hline 
\end{tabular}}
\end{center}
\caption {Fidelity estimators for our 5-qubit experiments with IBM's 7-qubit Jakarta quantum computer. 
} 
\label {t:jakarta}
\end {table}
\subsection {Additional tables: Google's experimental data and simulations} 
\label {s:app.data.g}

 

Table \ref {t:google-experiment} 
gives the fidelity estimators for the ten circuits of the Google quantum supremacy experiment for $n=12,14$. It can be compared Table \ref {t:QVM} which gives the same information for simulations of the same circuits by the Google's Weber QVM simulator.

{\bf Remark:} The disagreements of Formula (77) between the Google experimental data and our ``Weber QVM" simulations is based on using the average fidelity values for the simulation and the individual fidelity values for the Google experimental data. (Using the average fidelity values is considerably simpler and sufficient for our analysis, and, in addition, the Google team has not provided the individual fidelities used in Formula (77).) 

Table \ref {t:dataFig6} gives the data for Figure \ref {fig:fid-readout}.
In Section 6 of \cite {RSK22} we presented two estimators for $\phi_{ro}$ and they are also described in Section \ref{s:fourier-readout}. One estimator is analogous to our fidelity estimator $U$ (=${\cal F}_{XEB}$) and the other was based on MLE. Expressing these estimators using Fourier--Walsh coefficients and using fast Fourier--Walsh transform allows us to compute these estimators for $12 \le n \le 30$ and the outcomes are given in Table \ref {t:phi-rho-comp}. The MLE method for estimating $\phi_{ro}$ gives lower variance over ten circuits and it appears that part of the variance for the ${\cal F}_{XEB}$-style estimator for $\phi_{ro}$ reflects the instability of the predictor and another part reflects the different behavior of different circuits. Also here, the coefficient of variation for the MLE-based estimators for $\phi_{ro}$ is lower for the experimental data compared to the simulation. 


\begin{table}

\begin{center}\resizebox{1.\textwidth}{!}{
 \begin{tabular}{||c c c c c c c c c c||}
 \hline
 $n$&File&(77)&$U$&$V$&MLE&$T$&$S$&alt-$\phi$&$\phi_{ro}$\\
 \hline\hline 
12&0&0.386&0.381&0.374&0.375&0.476&0.471&0.415&0.261\\
12&1&0.386&0.379&0.373&0.372&0.476&0.472&0.240&0.151\\
12&2&0.386&0.350&0.356&0.359&0.457&0.462&0.242&0.152\\
12&3&0.386&0.382&0.379&0.376&0.473&0.470&0.411&0.258\\
12&4&0.386&0.384&0.369&0.371&0.479&0.470&0.257&0.161\\
12&5&0.386&0.381&0.368&0.369&0.475&0.467&0.400&0.252\\
12&6&0.386&0.359&0.366&0.367&0.462&0.467&0.292&0.184\\
12&7&0.386&0.382&0.353&0.360&0.468&0.450&0.263&0.165\\
12&8&0.386&0.351&0.371&0.367&0.466&0.479&0.341&0.214\\
12&9&0.386&0.353&0.362&0.370&0.457&0.462&0.253&0.159\\
12&\textbf{Avg}&&\textbf{0.370}&\textbf{0.367}&\textbf{0.369}&\textbf{0.469}&\textbf{0.467}&\textbf{0.311}&\textbf{0.196}\\
12&\textbf{Std}&&\textbf{0.014}&\textbf{0.008}&\textbf{0.005}&\textbf{0.008}&\textbf{0.007}&\textbf{0.066}&\textbf{0.042}\\
\hline
\multicolumn{8}{c}{} \\ \hline
 $n$&File&(77)&$U$&$V$&MLE&$T$&$S$&alt-$\phi$&$\phi_{ro}$\\
 \hline\hline 
14&0&0.332&0.336&0.326&0.327&0.439&0.432&0.335&0.195\\
14&1&0.332&0.332&0.330&0.328&0.439&0.438&0.402&0.234\\
14&2&0.332&0.328&0.329&0.326&0.441&0.442&0.338&0.197\\
14&3&0.332&0.334&0.332&0.331&0.435&0.433&0.237&0.138\\
14&4&0.332&0.322&0.327&0.326&0.439&0.442&0.463&0.269\\
14&5&0.332&0.330&0.328&0.326&0.439&0.437&0.252&0.146\\
14&6&0.332&0.332&0.324&0.327&0.437&0.431&0.317&0.184\\
14&7&0.332&0.329&0.328&0.328&0.441&0.440&0.306&0.178\\
14&8&0.332&0.325&0.324&0.325&0.442&0.441&0.375&0.218\\
14&9&0.332&0.330&0.331&0.331&0.442&0.443&0.363&0.211\\
14&\textbf{Avg}&&\textbf{0.330}&\textbf{0.328}&\textbf{0.328}&\textbf{0.439}&\textbf{0.438}&\textbf{0.339}&\textbf{0.197}\\
14&\textbf{Std}&&\textbf{0.004}&\textbf{0.003}&\textbf{0.002}&\textbf{0.002}&\textbf{0.004}&\textbf{0.061}&\textbf{0.036}\\

    \hline
\end{tabular}}
\caption {The various estimators for $\phi$ for the the 10 files with $n=12,14$ for Google 2019 experimental data. (Compare it to Table \ref {t:QVM} for the QVM simulations.) Here, the  (77) values were reported in \cite {Aru+19} and are based on individual qubit- and gate- fidelity values from the Google 2019 experiment. The fidelity estimators $U$, $V$ and $MLE$  agree on average very well with the (77) predictions. (This was our second concern in \cite {KRS23}.) Here we use the ${\cal F}_{XEB}$-style estimator for $\phi_{ro}$; Using the MLE estimation for $\phi_{ro}$ gives similar averaged values and smaller deviations (Table \ref {t:mle-phiro-sim}).
} 
\label{t:google-experiment}
\end{center}
\end{table}

\newpage
\begin{table}[H]
\begin{center}\resizebox{1.0\textwidth}{!}{
 \begin{tabular}{||c c c c c c c c c c c||} 
 \hline
 model/n&12&14&16&18&20&22&24&26&28&30\\ [0.5ex] 

 \hline\hline 
    alt-$\phi$&0.3305&0.2736&0.2145&0.1881&0.1637&0.1351&0.1089&0.0898&0.0737&0.0639\\
    MLE&0.3687&0.3275&0.2725&0.2444&0.2184&0.1651&0.1407&0.1140&0.0948&0.0823\\
     \hline 
\end{tabular}}
\caption {Estimating $\phi$ (alt-$\phi$) using average $\phi_{ro}$ estimated from the 10 files for $n=12 -30$ with Fourier analysis. (Here, we used the MLE version of the $\phi_{ro}$ estimator.) This data is presented in Fig. \ref{fig:fid-readout}. }
\label {t:dataFig6}
\end{center}
\end {table}

\begin{table}[H]
{\begin{center} \textbf{{$\phi_{ro}$ Estimated using MLE}} \end{center}}

\begin{center}\resizebox{0.95\textwidth}{!}{
 \begin{tabular}{||c c c c c c c c c c c||} 
 \hline
 File/$n$&12&14&16&18&20&22&24&26&28&30\\ [0.5ex] 
 \hline\hline 
0&0.222&0.190&0.198&0.211&0.177&0.195&0.184&0.186&0.148&0.148\\
1&0.187&0.225&0.177&0.184&0.200&0.180&0.154&0.154&0.152&0.118\\
2&0.164&0.189&0.187&0.189&0.168&0.164&0.164&0.148&0.156&0.118\\
3&0.196&0.183&0.196&0.187&0.224&0.194&0.160&0.165&0.151&0.153\\
4&0.196&0.207&0.172&0.196&0.197&0.177&0.175&0.147&0.151&0.151\\
5&0.253&0.149&0.175&0.163&0.190&0.181&0.171&0.153&0.136&0.166\\
6&0.204&0.169&0.192&0.204&0.201&0.181&0.161&0.144&0.163&0.137\\
7&0.166&0.174&0.179&0.192&0.185&0.171&0.168&0.136&0.126&0.132\\
8&0.194&0.199&0.176&0.172&0.190&0.188&0.169&0.176&0.143&0.14\\
9&0.178&0.206&0.181&0.165&0.181&0.179&0.164&0.160&0.119&0.138\\
\textbf{Avg}&\textbf{0.196}&\textbf{0.189}&\textbf{0.183}&\textbf{0.186}&\textbf{0.191}&\textbf{0.181}&\textbf{0.167}&\textbf{0.157}&\textbf{0.144}&\textbf{0.140}\\
\textbf{Std}&\textbf{0.025}&\textbf{0.021}&\textbf{0.009}&\textbf{0.015}&\textbf{0.015}&\textbf{0.009}&\textbf{0.008}&\textbf{0.014}&\textbf{0.013}&\textbf{0.014}\\
    \hline
\end{tabular}}
\end{center}

{\begin{center} \textbf{$\phi_{ro}$ Estimated using U-type estimator} \end{center}}

\begin{center}\resizebox{0.95\textwidth}{!}{
 \begin{tabular}{||c c c c c c c c c c c||} 
 \hline
 File/$n$&12&14&16&18&20&22&24&26&28&30 \\ [0.5ex] 
 \hline\hline 
0&0.261&0.195&0.181&0.219&0.178&0.193&0.185&0.186&0.148&0.146\\
1&0.151&0.234&0.176&0.192&0.201&0.178&0.152&0.152&0.151&0.118\\
2&0.152&0.197&0.185&0.195&0.170&0.170&0.163&0.147&0.156&0.118\\
3&0.258&0.138&0.205&0.188&0.222&0.196&0.160&0.165&0.151&0.154\\
4&0.161&0.269&0.174&0.198&0.197&0.180&0.177&0.147&0.151&0.150\\
5&0.252&0.146&0.202&0.165&0.185&0.181&0.174&0.153&0.136&0.166\\
6&0.184&0.184&0.187&0.202&0.201&0.178&0.159&0.141&0.162&0.139\\
7&0.165&0.178&0.176&0.196&0.183&0.170&0.169&0.135&0.126&0.131\\
8&0.214&0.218&0.176&0.180&0.192&0.189&0.170&0.174&0.144&0.140\\
9&0.159&0.211&0.181&0.162&0.186&0.182&0.163&0.161&0.119&0.140\\
\textbf{Avg}&\textbf{0.196}&\textbf{0.197}&\textbf{0.184}&\textbf{0.190}&\textbf{0.192}&\textbf{0.182}&\textbf{0.167}&\textbf{0.156}&\textbf{0.144}&\textbf{0.140}\\
\textbf{Std}&\textbf{0.044}&\textbf{0.037}&\textbf{0.010}&\textbf{0.016}&\textbf{0.014}&\textbf{0.008}&\textbf{0.009}&\textbf{0.015}&\textbf{0.013}&\textbf{0.014}\\
    \hline
\end{tabular}}
\end{center}
\caption{Comparing two estimators for $\phi_{ro}$ from \cite {RSK22} on the Google experimental data. The top table uses MLE and the bottom table uses the ${\cal F}_{XEB}$-type estimator.
Like for the estimators for the fidelity studied in \cite {RSK22}, for small number of qubits, the variance over the ten experimental circuits is smaller for the MLE-estimator.
The coefficient of variation of the MLE estimator for $\phi_{ro}$ is also somewhat smaller for the experimental data compared to QVM-simulation, see Table \ref {t:mle-phiro-sim}. }
\label {t:phi-rho-comp}
\end {table}

\begin{table}[ht]
\centering
    \begin{tabular}[t]{||c c c c||}    
    \hline
    n&File&$\phi$&$\phi_{ro}$\\[0.5ex] 
    \hline\hline 
     12&0&0.332&0.326\\
        12&1&0.322&0.271\\
        12&2&0.326&0.257\\
        12&3&0.317&0.207\\
        12&4&0.327&0.301\\
        12&5&0.332&0.329\\
        12&6&0.317&0.189\\
        12&7&0.322&0.246\\
        12&8&0.326&0.249\\
        12&9&0.324&0.229\\
    \textbf{12}&\textbf{Avg}&\textbf{0.325}&\textbf{0.261}\\
    \textbf{12}&\textbf{Std}&\textbf{0.005}&\textbf{0.045}\\
    \hline 
    \end{tabular}\hfill%
    \begin{tabular}[t]{||c c c c||} 
    \hline
    n&File&$\phi$&$\phi_{ro}$\\[0.5ex] 
    \hline\hline 
    14&0&0.269&0.232\\
14&1&0.270&0.254\\
14&2&0.266&0.210\\
14&3&0.270&0.276\\
14&4&0.271&0.265\\
14&5&0.266&0.240\\
14&6&0.266&0.245\\
14&7&0.262&0.172\\
14&8&0.274&0.289\\
14&9&0.266&0.235\\
    \textbf{14}&\textbf{Avg}&\textbf{0.268}&\textbf{0.242}\\
    \textbf{14}&\textbf{Std}&\textbf{0.003}&\textbf{0.032}\\
    \hline 
    \end{tabular}
\caption{Estimating the values of the pair ($\phi$, $\phi_{ro}$) with $q=0.038$ (see Section \ref {s:mle-phiro}) for simulation with Google’s full noise model of the Google circuits for 12 and 14 qubits.}
\label {t:mle-phiro-sim}
\end{table}

\subsection {Instability of large and small Fourier degrees}
\label {s:stab}
We give here a more detailed data for the estimator $\Lambda_k$ of degree $k$- Fourier contribution of $n$-qubit circuits were we indicate also the values for $k=1,n-1,n$ 
which are rather unstable. This information is related to a comment by Carsten Voelkmann to an early version to the paper; it is less relevant to our analysis throughout the paper but is of some interest on its own and also in comparison to the finding of Section \ref {s:dolev} and especially Figure \ref {fig:Dolev3}.

\begin{figure}
\centering
\includegraphics[width=1.0\textwidth]{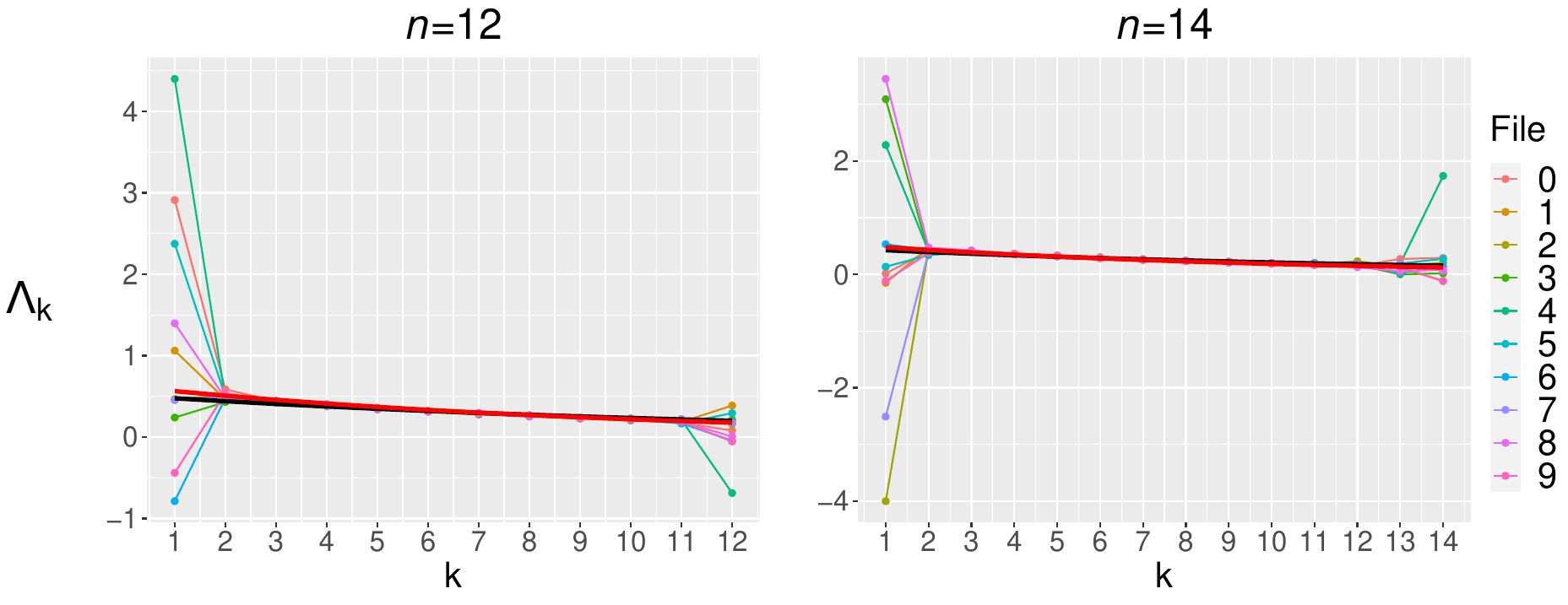}
\caption{The decay of the relative contribution of degree-$k$ Fourier coefficient degree $k$ Fourier coefficients for $n=12,14$-qubit random circuits on Weber QVM simulators with Google's full noise model. We added the values for $k=1,11,12$. The values are rather spread out for $k=n$, and very spread out for $k=1$. 
}
\label{fig:12gs}
\end{figure}

\begin{figure}[h]
\centering
\includegraphics[width=\textwidth]{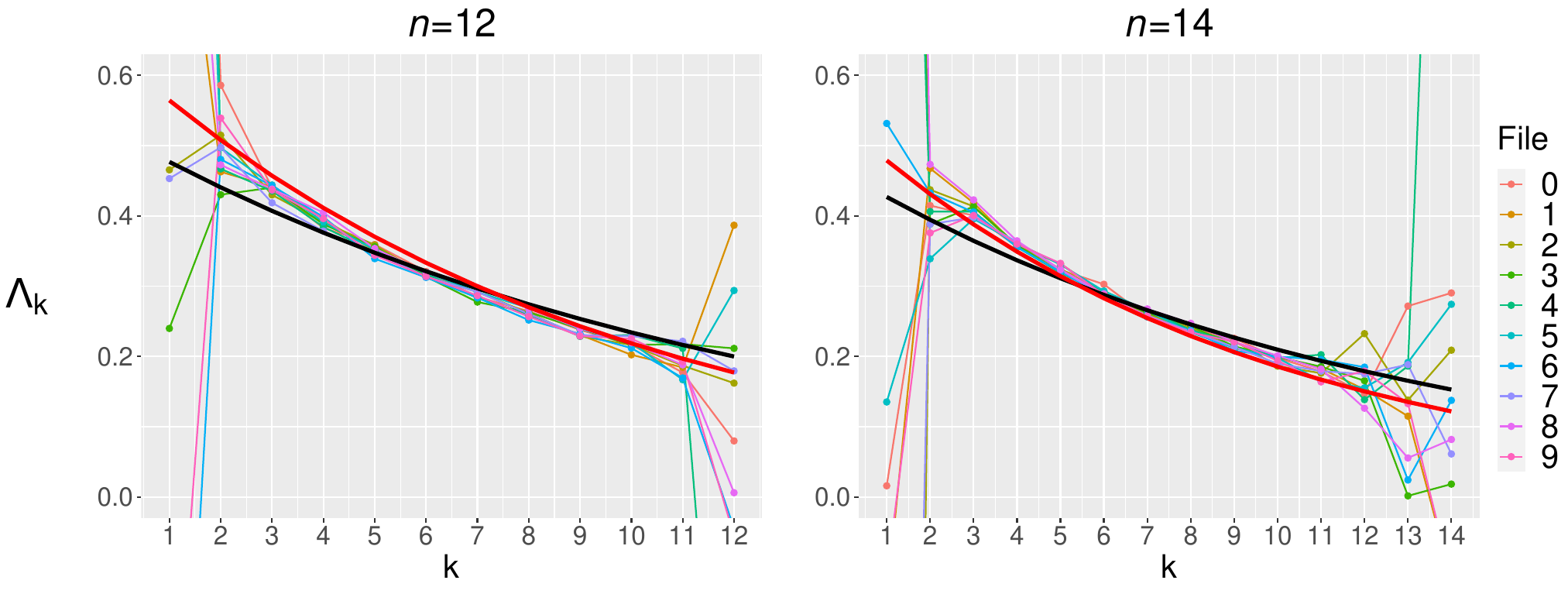}
\caption{The decay of the relative contribution of degree-$k$ Fourier coefficient degree $k$ Fourier coefficients for $n$-qubit random circuits, $n=12,14$ on Weber QVM simulators with Google's full noise model. We indicated the values for $k=1,n-1,n$ via broken line segments. } 
\label{fig:ohad12+14F}
\end{figure}

Figure \ref {fig:12gs} shows the relative contribution of degree-$k$ Fourier coefficient  for 12-qubit and 14-qubit random circuits on Weber QVM simulators with Google's full noise model. Figure \ref {fig:ohad12+14F} presents the same data in a different manner: we show the data for $n=12,14$ so that the contributions for the range $2 \le k \le n-1$ is clearly presented and yet the behavior for $k=1,n-1,n$ is indicated by broken lines. Overall the behavior for the extreme values is very unstable both for the experimental data and for simulation. (See Section \ref {s:tables} for the full data.) IBM's Fake Guadalupe simulations (Figure \ref{fig:Fourier-fake-guadalupe}) also exhibit unstable behavior of the extreme  values of $\Lambda_k$ but to a lesser amount compared to both the Google's simulation and the experimental data.  


\comment{
\begin{figure}[h]
\centering
\includegraphics[width=\textwidth]{google12+14exp}
\caption{The decay of degree $k$ Fourier contribution for Google’s experimental  samples for Google circuits for 12 and 14 qubits. The decrease in the Fourier contribution is consistent (in fact a little smaller) with the physical readout errors and it appears that there is no additional effect of a similar nature of the gate errors. (See also Figure \ref {fig:Fourier_Delta_Google_data} for the average begavior for values of $n$ between 12 and 26.) The solid red line is based on $s=$, $q=$,  [to be completed]. We indicated the values for $k=1,n-1,n$ via broken line segments. This Figure can be compared to 
Figure \ref {fig:ohad12+14F} for Google's simulation. 
}
\label{fig:ohad12+14exp}
\end{figure}
}

\newpage
\subsection {Data from a Harvard/MIT/QuEra experiment}
\label {s:dolev}

The 2024 paper by Bluvstein et al. \cite {Blu+23} describes a logical quantum processor based on reconfigurable neutral atom arrays. The experiment was conducted by a group from Harvard, MIT, and the quantum computing company QuEra, and was led by Michael Lukin.
We applied our analysis to one of the experiments described in the paper where a pair of logical IQP states on 12 logical qubits were simultaneously created based on  [[8,3,2]] color code. Altogether, the experiment described a processor of 64 physical qubits acting simultaneously on two arrays of 32 qubits. Transversal logical gates were applied on the physical qubits that describe the logical qubits of the encoded state. Once the physical qubits were measured a post-selection process with 16 threshold values based on error-detection was applied. A normalized ${\cal F}_{XEB}$ fidelity-estimator (which is our estimator $V$) was applied on the measured logical qubits after the post-selection stage.\footnote {The size of the full sample was 188k, and there was a pre-selection process based on physical criteria.} 

\begin{figure}[h]
\centering
\includegraphics[width=\textwidth]{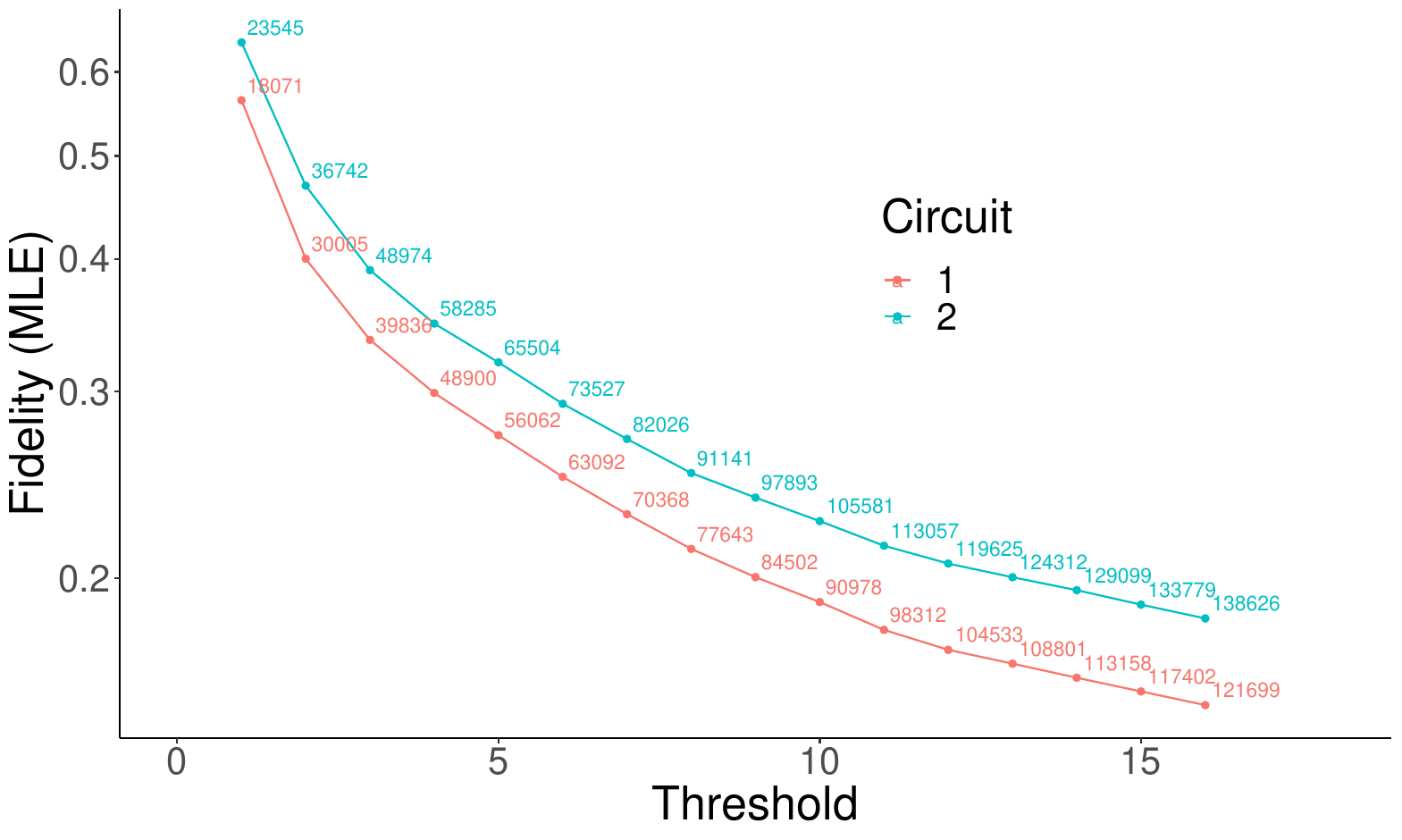}
\caption{The dependence of the MLE estimated fidelity for the samples for the two 12 logical qubit IQP states  as a function of the post-selection threshold. The physical circuits are based on natural atoms, see \cite {Blu+23}.} 
\label{fig:Dolev2}
\end{figure}

\begin{figure}
\centering
\includegraphics[width=\textwidth]{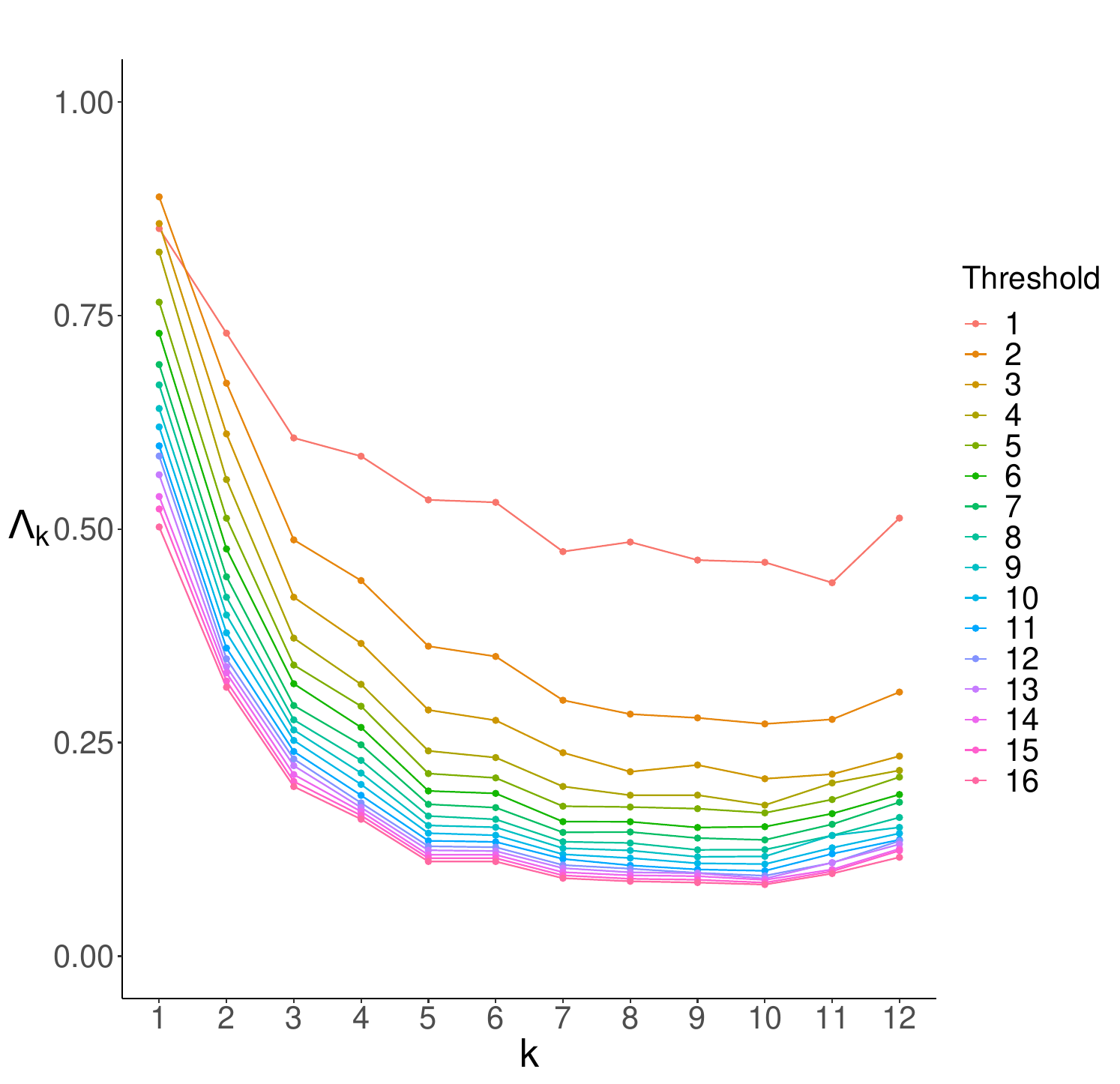}
\caption{The decay of the relative contribution of degree-$k$ Fourier coefficient  for one  12-qubit logical IQP circuits from \cite {Blu+23} as a function of the post-selection threshold. (This is circuit 2 in Table \ref {tab:Dolev1}; it has somewhat higher fidelity compared to the other circuit that the authors attribute to a slightly more successful calibration.) The behavior for the other 12-qubit IQP circuit is similar, see Figure \ref {fig:Dolev7}. The stability of $\Lambda_k$ along the different thresholds and the different values of $k$ is notably different than the behavior we witnessed for other samples from experiments and simulations discussed in this paper.}
\label{fig:Dolev3}
\end{figure}

We applied our detailed analysis on these samples 
and the results are described in Table \ref {tab:Dolev1}, Figure \ref {fig:Dolev2}, and Figure \ref {fig:Dolev3}. We were pleased to see that our programs applied smoothly and gave relevant analysis. As seen from Table \ref {tab:Dolev1} the fidelity estimators $V$, MLE, and
$S$ agree well with each other. The fact that the estimator $S$ agrees well with the normalized fidelity measures $V$ and MLE means that contrary to the samples from the Google quantum supremacy experiment the empirical variance agrees with what basic noise models, such as the Google noise model gives; this deserves a further study.  
The values for $V$ that we computed agree with those reported in \cite {Blu+23}. 

\begin{figure}
\centering
\includegraphics[width=\textwidth]{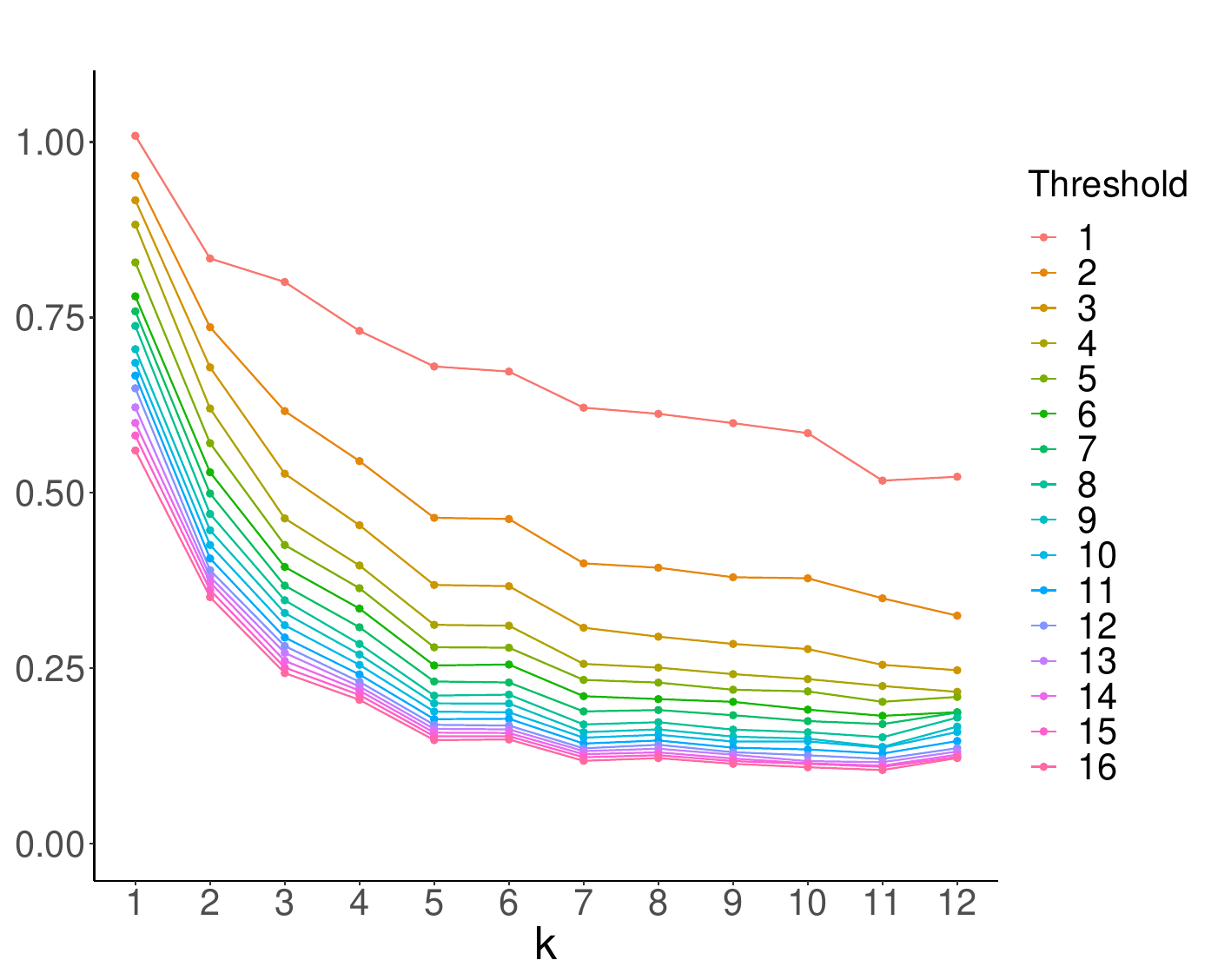}
\caption{The decay of the relative contribution of degree-$k$ Fourier coefficient  for the other  12-qubit logical IQP circuits from \cite {Blu+23} as a function of the post-selection threshold.}
\label{fig:Dolev7}
\end{figure}

The effective readout error $q$ is close to 0.1 for
the largest 13 postselected samples, and then decreases 
for the remaining three samples reaching the value  $q=0.025$ for the last sample. 
Figures \ref {fig:Dolev3} and \ref {fig:Dolev7} describe the estimators $\Lambda_k$ for degree-$k$ Fourier contributions for the two IQP circuits. (The graphs appear to reach a plateaux for $k \ge 7$, so they appear not to fit well the behavior of our two-parameter model.)   These $\Lambda_k$-estimates are stable between different thresholds and for the entire range of values of $k$, $1 \le k \le 12$. This appears to be a different behavior than that of other samples that we considered throughout the paper. It would be interesting 
to compare all the 4,095 Fourier--Walsh coefficients of the empirical samples (say, for the largest samples that do not represent post selection) with that of the ideal probabilities. 

{\bf Remark:} 
The paper \cite {Blu+23} contains quite a few other interesting experiments on the neutral atom quantum computer. A large number of error correcting codes including surface codes with distances 3,5, and 7, color codes and high dimensional codes were created; to check that the state they created is indeed what they aimed for, the experimentalists measured the state against a large number of product states. Several logical states such as GHZ state (with 4 logical qubits)  and IQP state with various number of logical qubits were created. 

Sharing the the raw data for the many experiments of \cite {Blu+23} 
is important for checking various aspects of the experiments. It is also important for further analysis of the nature of noise, both for logical states and for corresponding physical states. We note that \cite {Blu+23} already reports on both ``logical" sampling experiments, and on ``physical" sampling experiments such as an analysis of entanglement entropy on the physical level.

\begin{table}[H]\begin{center}
 \begin{tabular}{||c c c c c c c c c c||} 
 \hline
i&j&N&U&V&MLE&T&S&$s$&$q$\\[0.5ex] 
\hline\hline 
1&1&18,071&4.397&0.546&0.564&1.58&0.557&0.761&0.027\\
1&2&30,005&3.104&0.386&0.400&1.208&0.426&0.809&0.066\\
1&3&39,836&2.551&0.317&0.336&1.018&0.359&0.819&0.085\\
1&4&48,900&2.215&0.275&0.299&0.904&0.319&0.807&0.095\\
1&5&56,062&2.017&0.251&0.273&0.825&0.291&0.760&0.098\\
1&6&63,092&1.846&0.229&0.249&0.760&0.268&0.734&0.103\\
1&7&70,368&1.703&0.212&0.230&0.703&0.248&0.700&0.106\\
1&8&77,643&1.581&0.196&0.213&0.657&0.231&0.689&0.112\\
1&9&84,502&1.490&0.185&0.201&0.623&0.220&0.668&0.115\\
1&10&90,978&1.403&0.174&0.190&0.590&0.208&0.646&0.117\\
1&11&98,312&1.323&0.164&0.179&0.559&0.197&0.628&0.120\\
1&12&104,533&1.266&0.157&0.171&0.537&0.189&0.618&0.123\\
1&13&108,801&1.225&0.152&0.166&0.519&0.183&0.600&0.123\\
1&14&113,158&1.183&0.147&0.161&0.503&0.177&0.583&0.123\\
1&15&117,402&1.144&0.142&0.157&0.486&0.171&0.569&0.123\\
1&16&121,699&1.109&0.138&0.152&0.472&0.166&0.552&0.123\\
\hline
\multicolumn{8}{c}{} \\ \hline
i&j&N&U&V&MLE&T&S&$s$&$q$\\[0.5ex] 
\hline\hline 
2&1&23,545&5.526&0.686&0.640&1.983&0.699&0.809&0.021\\
2&2&36,742&3.938&0.489&0.469&1.471&0.518&0.805&0.050\\
2&3&48,974&3.206&0.398&0.390&1.225&0.432&0.835&0.071\\
2&4&58,285&2.777&0.345&0.348&1.079&0.380&0.818&0.081\\
2&5&65,504&2.534&0.315&0.320&0.987&0.348&0.770&0.083\\
2&6&73,527&2.319&0.288&0.292&0.904&0.318&0.739&0.087\\
2&7&82,026&2.133&0.265&0.271&0.840&0.296&0.726&0.093\\
2&8&91,141&1.972&0.245&0.251&0.784&0.276&0.710&0.098\\
2&9&97,893&1.863&0.231&0.238&0.744&0.262&0.680&0.099\\
2&10&105,581&1.765&0.219&0.227&0.707&0.249&0.662&0.102\\
2&11&113,057&1.674&0.208&0.215&0.675&0.238&0.643&0.104\\
2&12&119,625&1.599&0.199&0.207&0.646&0.228&0.623&0.105\\
2&13&124,312&1.547&0.192&0.201&0.626&0.220&0.604&0.104\\
2&14&129,099&1.499&0.186&0.195&0.607&0.214&0.590&0.105\\
2&15&133,779&1.453&0.181&0.189&0.588&0.207&0.574&0.105\\
2&16&138,626&1.406&0.175&0.183&0.569&0.201&0.556&0.105\\

\hline 
\end{tabular}
\caption {The various estimators for the pair of 16 files for IQP states with 12 qubits. For the noiseless circuit $U=8.05078 $. Here $i=1,2$ is the index of the circuit, $j=1,2,\dots,16$ indicates the post selection threshold, and $N$ is the size of the post-selected circuit. The fidelity estimators $V$, $MLE$, and $S$ agree well with each other.
The effective readout error $q$ is close to 0.1 for the 13 largest post selected samples, and then gradually decreases to 0.025 for the strictest post selection. 
} 
\label{tab:Dolev1}
\end{center}
\end{table}

\newpage
\section{Appendix B: fidelity estimators}
\label{app:est}

Here we give the full formulas for the fidelity estimators ${\cal F}_{XEB}$ (=$U$), $V$, MLE, $T$ and $S$, that we use throughout the paper. For further details, including motivation, properties, bias, etc. we refer to \cite{RSK22}.



Recall that $M=2^n$. Let $w_1,...,w_{2^n}$ denote the probabilities over all $\{0,1\}^n$ bitstrings, such that $\sum_{i=1}^{2^n} w_i=1$. Let ${\widetilde w}_i$ denotes the probability of the ith bitstring that was sampled according to the Google model, where we have $N=500,000$ such samples, ${\widetilde w}_1,...,{\widetilde w}_N$. We further denote the total appearances of the ith bitstring by $n_i$, such that $\sum_{i=1}^{2^n} n_i = N = 500,000$.

The XEB estimator, stated also in \eqref{e:fxeb}, in our notations is given by
\begin{equation}\label{eq:U}
U \coloneqq 2^n\frac {1}{N}\sum_{j=1}^N {\widetilde w}_{j}-1 = 2^n\frac {1}{N}\sum_{j=1}^{2^n} n_j w_j -1.
\end{equation}

In \cite{RSK22} we showed that this estimator is actually biased, and we suggested a variant of the estimator that accounts for the bias in each individual circuit. We denote $ w^{(k)}=\sum_{i=1}^M w_i^k$, and put forward the estimator
\begin{equation}\label{eq:V} 
V \coloneqq U/(M w^{(2)}-1).
\end{equation}

Another estimator we discussed is the maximum likelihood estimator. We derived the derivative of the log-likelihood, and thus, the MLE can be found by solving the following equation
\begin{equation}\label{eq:MLE}
MLE = \phi \text{ such that } \sum_{j=1}^N \frac{\widetilde w_{j}-1/M}{\phi \widetilde w_{j}+(1-\phi)/M}=0.
\end{equation}
There is no explicit solution, and a numerical method for solving this equation was suggested.

Another estimator we put forward was based on the second moment of the distribution of bitstrings. We defined
\begin{equation}\label{eq:T_squared}
T^2 = \frac{M(M+1)}{(N^2-N)(M-1)}\left(\sum_{i=1}^M n_i^2 -N-(N^2-N)/M\right)
\end{equation}
We have shown that $T^2$ is an unbiased estimator for $\phi^2$, taking expectation over all random circuits, and thus we suggested $\max(0,T)$ as a biased estimator for $\phi$.

Recall that the estimator $U$ is unbiased over all random circuits but it is biased over an individual random circuit. The estimator $V$ which is a normalized version of $U$ is unbiased also for a specific circuit. Similarly, $T^2$ is an unbiased estimator for $\phi^2$ (under the Google noise model) over all random circuits but it is biased for an individual circuit. 

Here we suggest an unbiased estimator $S^2$ which is a normalized version of $T^2$. We note that to compute $T^2$ we did not need to know any amplitude (probabilities), only the appearances. $S^2$ requires knowing all the amplitudes.

\begin{equation}\label{eq:S_squared}
    S^2 = \frac{\sum_{i=1}^M n_i^2 -N-(N^2-N)/M}{(N^2-N)(w^{(2)}-1/M)}
\end{equation}

{\bf Remark:} For the samples of the Google experiment as well as the Google QVM simulations ${\cal P}_C(x)$ behaves like a Porter--Thomas distribution and subsequently $U,V$ and MLE give rather similar values, and also $T$ and $S$ give similar values. The value of $T$ (or $S$) is substantially larger than the actual fidelity as estimated by MLE. For the IBM ``Fake Guadalupe" simulation (as well as our own 5-qubit samples) the ideal probability distributions are farther apart from a Porter--Thomas distribution, and (subsequently) $V$ and MLE give much more stable results than $U$ and similarly $S$ is much more stable than $T$. For the IBM ``Fake Guadalupe" simulation the value of $S$ is similar to the MLE value.

\newpage
\section {Appendix C: a list of recent NISQ experiments}
\label {s:nisqexp}
Here is a (partial) list of recent NISQ experiments that can be used for statistical analysis of samples from NISQ computers either based on tools of our papers or additional tools. Fourier methods may be relevant to several theoretical and empirical aspects related to these papers.

\begin {enumerate} 
\item
Random circuit sampling experiments: There are several related experiments on random circuit sampling. Most recently, the Google team presented a study \cite {Mor+23} which is similar to \cite {Aru+19} with larger circuits. Earlier a team from USTC repeated the Google RCS experiment with their Zuchongzhi quantum computer \cite {Wu+21}. 
\item
There are several works stating progress on quantum error correction for both on superconducting quantum computers, ion-trapped quantum computers, and other types of quantum computers: the paper \cite {Goo23} by the Google team describes realizations of distance-3 and distance-5 surface codes on Sycamore; Quantinuum's ion trap fault tolerance experiments \cite {Ryan+21,Ryan+22}; the Yale group's experiment \cite {Sivak+22}; a recent paper by Harvard/MIT/QuEra group \cite {Blu+23} (See Section \ref {s:dolev} for prelimimary analysis of data from that paper.) 
\item 
There are various interesting experiments by the Google AI team using the Sycamore quantum computer.  The paper \cite {MIQ+22} describes ``time crystals": The paper \cite {JZL+22} describes simulations of 9-qubit states related to models of wormholes.
\item 
Several experiments from the IBM team (and other teams working on IBM quantum computers) for demonstrating high quantum volume. See, for example, \cite {Jur+21}. 
\item 
A recent work  \cite {KEA+23} by the IBM quantum team on the Ising model using a 127-qubit computation demonstrating various techniques of error mitigation. 
\item 
Two experiments by Pokharel and Lidar \cite {PokLid23,PokLid22} stating speedup of Bernstein-Vazirani's and Grover's algorithms on IBM's quantum computers.   
\item
In the paper \cite {PBE22} by a team from Los Alamos National Laboratory, the authors performed their own series of quantum volume calculations on twenty four NISQ devices offered by IBM, IonQ, Rigetti, Oxford Quantum Circuits, and Quantinuum (formerly Honeywell).
\item 
Majorana zero-mode and Kitaev's chain based on (ordinary) quantum computers. There is much effort (largely encouraged by Microsoft) to demonstrate Majorana zero modes in physical systems and in parallel there is also some effort to simulate such quantum states on quantum computers. See \cite {SBEP21} (on an IBM quantum computer), \cite {Ran22} (on an IBM quantum computer) and \cite {Mi+22} (on Google's quantum computer).  
\item 
Boson sampling experiments. We already mentioned the experiment described in \cite {Zho+20} by a group from USTC. 
Another experiment by a team from Xanadu is described in \cite {MLA+22}. 
\end {enumerate}

\section{Tables for Figures 3,4,5,7,11, and 12}
\label {s:tables}
\subsection{Average Fourier contribution for the experimental data (Figure 3)}

\begin{table}[H]\begin{center}
\begin{adjustbox}{max width=\textwidth}
 \begin{tabular}{||c c c c c c c || c ||c c c c c c c||} 
 \hline
\textbf{K/n}&\textbf{12}&\textbf{16}&\textbf{18}&\textbf{22}&\textbf{24}&\textbf{26}&&\textbf{K/n}&\textbf{12}&\textbf{16}&\textbf{18}&\textbf{22}&\textbf{24}&\textbf{26}\\
\hline\hline 
\textbf{1}&0.875&1.601&0.372&2.786&-0.450&2.391&&\textbf{14}&-&0.203&0.181&0.135&0.123&0.104\\
\textbf{2}&0.506&0.408&0.422&0.502&-0.557&-0.245&&\textbf{15}&-&0.164&0.172&0.130&0.116&0.100\\
\textbf{3}&0.456&0.384&0.371&0.310&0.299&0.300&&\textbf{16}&-&0.276&0.194&0.119&0.110&0.096\\
\textbf{4}&0.419&0.346&0.335&0.275&0.230&0.227&&\textbf{17}&-&-&0.146&0.118&0.105&0.087\\
\textbf{5}&0.381&0.325&0.309&0.251&0.228&0.208&&\textbf{18}&-&-&-2.863&0.105&0.103&0.083\\
\textbf{6}&0.363&0.304&0.289&0.231&0.200&0.185&&\textbf{19}&-&-&-&0.066&0.092&0.081\\
\textbf{7}&0.338&0.285&0.272&0.211&0.194&0.171&&\textbf{20}&-&-&-&0.111&0.087&0.075\\
\textbf{8}&0.325&0.269&0.256&0.198&0.179&0.153&&\textbf{21}&-&-&-&-0.305&0.067&0.088\\
\textbf{9}&0.302&0.253&0.241&0.185&0.168&0.144&&\textbf{22}&-&-&-&2.634&0.059&0.034\\
\textbf{10}&0.269&0.24&0.227&0.172&0.157&0.135&&\textbf{23}&-&-&-&-&-0.577&0.032\\
\textbf{11}&0.262&0.229&0.215&0.162&0.145&0.128&&\textbf{24}&-&-&-&-&14.826&-0.503\\
\textbf{12}&0.418&0.215&0.204&0.152&0.140&0.119&&\textbf{25}&-&-&-&-&-&-0.180\\
\textbf{13}&-&0.197&0.191&0.145&0.130&0.113&&\textbf{26}&-&-&-&-&-&-3.246\\
\hline
\end{tabular}
\end{adjustbox}
\caption {The average over ten circuits of $\Lambda_k$ for the experimental samples for $n=12,16,18,22,24,26$.} 
\label{tab:Figure3}
\end{center}
\end{table}

\subsection{Fourier contribution for Google data}

\begin{table}[H]\begin{center}
 \begin{tabular}{||c c c c c c c c c c c||} 
 \hline
K/File&0&1&2&3&4&5&6&7&8&9\\[0.5ex] 
\hline\hline 
1&1.923&0.685&0.462&0.804&1.904&1.492&0.191&-0.006&1.105&0.185\\
2&0.552&0.524&0.505&0.503&0.476&0.48&0.630&0.460&0.484&0.444\\
3&0.416&0.452&0.445&0.476&0.473&0.472&0.469&0.447&0.468&0.447\\
4&0.420&0.423&0.391&0.422&0.414&0.435&0.424&0.413&0.413&0.432\\
5&0.393&0.391&0.379&0.387&0.391&0.384&0.384&0.363&0.364&0.377\\
6&0.378&0.376&0.341&0.368&0.365&0.360&0.361&0.345&0.378&0.36\\
7&0.339&0.337&0.344&0.344&0.329&0.330&0.342&0.342&0.356&0.323\\
8&0.326&0.338&0.312&0.353&0.334&0.309&0.309&0.314&0.318&0.341\\
9&0.295&0.300&0.298&0.312&0.317&0.329&0.290&0.302&0.304&0.268\\
10&0.267&0.281&0.311&0.295&0.288&0.219&0.271&0.216&0.262&0.283\\
11&0.237&0.377&0.265&0.241&0.286&0.311&0.294&0.170&0.171&0.269\\
12&0.195&1.796&0.091&0.210&1.552&0.449&0.261&-0.658&0.312&-0.023\\
\hline
\multicolumn{11}{c}{} \\ 
\hline 
K/File&0&1&2&3&4&5&6&7&8&9\\[0.5ex] 
\hline\hline
1&0.269&1.698&-0.279&0.482&1.043&-0.22&-0.099&-0.340&1.594&0.744\\
2&0.546&0.569&0.578&0.520&0.495&0.400&0.573&0.541&0.386&0.532\\
3&0.462&0.451&0.428&0.442&0.451&0.417&0.419&0.440&0.434&0.425\\
4&0.389&0.392&0.407&0.389&0.403&0.395&0.395&0.405&0.409&0.409\\
5&0.358&0.380&0.357&0.365&0.361&0.362&0.353&0.366&0.367&0.371\\
6&0.344&0.341&0.356&0.357&0.344&0.349&0.345&0.344&0.336&0.345\\
7&0.312&0.321&0.318&0.328&0.320&0.324&0.318&0.330&0.311&0.328\\
8&0.314&0.304&0.301&0.311&0.301&0.310&0.304&0.303&0.305&0.306\\
9&0.289&0.291&0.292&0.293&0.273&0.282&0.288&0.292&0.277&0.293\\
10&0.271&0.262&0.277&0.285&0.280&0.280&0.268&0.262&0.281&0.269\\
11&0.238&0.286&0.254&0.263&0.276&0.287&0.258&0.258&0.256&0.246\\
12&0.265&0.206&0.291&0.267&0.233&0.284&0.272&0.222&0.297&0.235\\
13&0.283&0.335&0.312&0.235&0.328&0.092&0.037&0.329&0.236&0.193\\
14&0.534&0.034&-0.002&-1.606&0.878&-0.556&0.416&-0.029&0.235&1.000\\
\hline 
\end{tabular}
\caption {The values of $\Lambda_k$ for the Google data for the ten experimental circuits for $n=12,14$.} 
\label{tab:Google_data_Fourier}
\end{center}
\end{table}

\subsection{Fourier contribution for Weber QVM simulations (Figure 4)}

\begin{table}[H]\begin{center}
 \begin{tabular}{||c c c c c c c c c c c||} 
 \hline
K/File&0&1&2&3&4&5&6&7&8&9\\[0.5ex] 
\hline\hline 
1&2.912&1.062&0.465&0.240&4.400&2.374&-0.787&0.453&1.397&-0.440\\
2&0.586&0.463&0.515&0.430&0.467&0.497&0.480&0.497&0.473&0.539\\
3&0.440&0.441&0.430&0.440&0.435&0.444&0.443&0.419&0.438&0.437\\
4&0.383&0.395&0.391&0.388&0.384&0.391&0.398&0.378&0.405&0.396\\
5&0.359&0.350&0.359&0.356&0.349&0.352&0.339&0.348&0.354&0.344\\
6&0.321&0.316&0.314&0.313&0.318&0.319&0.312&0.313&0.319&0.314\\
7&0.285&0.285&0.292&0.277&0.283&0.292&0.284&0.292&0.298&0.287\\
8&0.264&0.258&0.262&0.263&0.260&0.269&0.252&0.261&0.268&0.257\\
9&0.238&0.231&0.229&0.239&0.229&0.239&0.232&0.230&0.240&0.229\\
10&0.217&0.202&0.216&0.216&0.230&0.220&0.212&0.229&0.219&0.225\\
11&0.177&0.184&0.187&0.218&0.211&0.167&0.169&0.222&0.188&0.192\\
12&0.080&0.387&0.162&0.211&-0.688&0.294&-0.045&0.179&0.006&-0.056\\

\hline
\multicolumn{11}{c}{} \\ \hline
K/File&0&1&2&3&4&5&6&7&8&9\\[0.5ex] 
\hline\hline 
1&0.016&-0.149&-3.998&3.089&2.281&0.135&0.532&-2.508&3.448&-0.114\\
2&0.415&0.468&0.437&0.388&0.406&0.339&0.433&0.388&0.473&0.376\\
3&0.400&0.418&0.413&0.413&0.406&0.396&0.405&0.398&0.423&0.401\\
4&0.356&0.357&0.357&0.359&0.355&0.361&0.356&0.363&0.364&0.360\\
5&0.323&0.332&0.325&0.319&0.319&0.331&0.320&0.323&0.324&0.332\\
6&0.303&0.290&0.291&0.290&0.293&0.293&0.286&0.286&0.289&0.287\\
7&0.256&0.261&0.261&0.262&0.266&0.265&0.256&0.258&0.268&0.257\\
8&0.240&0.241&0.238&0.243&0.238&0.234&0.237&0.235&0.247&0.234\\
9&0.226&0.212&0.222&0.214&0.220&0.212&0.208&0.213&0.221&0.221\\
10&0.198&0.200&0.186&0.199&0.198&0.198&0.200&0.190&0.201&0.194\\
11&0.179&0.183&0.176&0.185&0.202&0.165&0.195&0.179&0.182&0.164\\
12&0.147&0.153&0.232&0.165&0.138&0.155&0.185&0.175&0.126&0.179\\
13&0.272&0.115&0.138&0.002&0.186&0.191&0.024&0.188&0.056&0.133\\
14&0.290&-0.117&0.209&0.018&1.737&0.274&0.137&0.061&0.082&-0.118\\

\hline 
\end{tabular}
\caption {The values of $\Lambda_k$ for the Weber QVM simulations for the ten experimental circuits for $n=12,14$.} 
\label{tab:Figure4}
\end{center}
\end{table}

\subsection{Simulations based on depolarizing noise on 1-gates (Figure 5)}

\begin{table}[H]\begin{center}
 \begin{tabular}{||c c c c c c c c c c c||} 
 \hline
K/File&0&1&2&3&4&5&6&7&8&9\\[0.5ex] 
\hline\hline 
1&0.431&0.467&0.434&0.469&0.495&0.459&0.468&0.472&0.445&0.533\\
2&0.431&0.418&0.419&0.434&0.418&0.413&0.445&0.428&0.410&0.411\\
3&0.390&0.402&0.388&0.393&0.398&0.391&0.394&0.404&0.395&0.390\\
4&0.375&0.382&0.385&0.374&0.372&0.381&0.372&0.379&0.379&0.374\\
5&0.365&0.364&0.359&0.367&0.364&0.364&0.37&0.357&0.361&0.358\\
6&0.360&0.354&0.349&0.363&0.354&0.350&0.353&0.356&0.352&0.352\\
7&0.350&0.346&0.351&0.351&0.349&0.348&0.345&0.348&0.347&0.339\\
8&0.339&0.345&0.342&0.351&0.346&0.347&0.333&0.341&0.342&0.342\\
9&0.339&0.341&0.324&0.331&0.327&0.341&0.327&0.336&0.336&0.338\\
10&0.345&0.326&0.332&0.349&0.326&0.346&0.326&0.336&0.309&0.343\\
11&0.291&0.302&0.347&0.333&0.274&0.354&0.34&0.310&0.288&0.330\\
12&0.208&0.071&0.273&0.250&-0.371&0.440&0.468&0.901&0.355&0.436\\
\hline
\multicolumn{11}{c}{} \\ \hline
K/File&0&1&2&3&4&5&6&7&8&9\\[0.5ex] 
\hline\hline 
1&0.455&0.44&0.443&0.481&0.422&0.310&0.397&0.416&0.434&0.554\\
2&0.363&0.379&0.353&0.373&0.374&0.337&0.376&0.363&0.381&0.361\\
3&0.345&0.351&0.335&0.365&0.367&0.34&0.350&0.363&0.347&0.348\\
4&0.334&0.332&0.343&0.330&0.323&0.329&0.326&0.336&0.327&0.338\\
5&0.317&0.322&0.318&0.32&0.313&0.320&0.313&0.322&0.315&0.316\\
6&0.307&0.309&0.309&0.313&0.309&0.310&0.308&0.306&0.311&0.306\\
7&0.293&0.298&0.298&0.301&0.307&0.294&0.300&0.302&0.303&0.307\\
8&0.294&0.295&0.29&0.299&0.288&0.291&0.292&0.296&0.293&0.294\\
9&0.292&0.280&0.289&0.293&0.288&0.284&0.289&0.288&0.290&0.283\\
10&0.288&0.290&0.278&0.290&0.295&0.284&0.280&0.282&0.278&0.279\\
11&0.273&0.283&0.282&0.275&0.287&0.274&0.278&0.271&0.278&0.270\\
12&0.263&0.269&0.308&0.295&0.301&0.287&0.299&0.262&0.280&0.289\\
13&0.215&0.346&0.442&0.299&0.270&0.219&0.265&0.306&0.265&0.288\\
14&0.281&-0.075&0.249&-0.534&0.546&0.139&0.450&0.773&0.145&0.389\\
\hline 
\end{tabular}
\caption {The values of $\Lambda_k$ for simulations based on depolarizing noise on 1-gates for the ten experimental circuits, $n=12,14$.} 
\label{tab:Figure5}
\end{center}
\end{table}

\subsection{Fake Guadalupe  simulations (Figure 7)}

\begin{table}[H]\begin{center}
 \begin{tabular}{||c c c c c c c c c c c||} 
 \hline
T/K&0&1&2&3&4&5&6&7&8&9\\[0.5ex] 
\hline\hline 
1&0.881&0.71&0.852&0.877&0.532&0.852&0.813&0.801&0.806&0.797\\
2&0.694&0.753&0.718&0.699&0.678&0.711&0.748&0.728&0.643&0.695\\
3&0.620&0.600&0.582&0.596&0.569&0.596&0.607&0.608&0.567&0.620\\
4&0.513&0.502&0.480&0.544&0.484&0.519&0.532&0.563&0.483&0.503\\
5&0.426&0.459&0.430&0.473&0.410&0.428&0.506&0.445&0.436&0.450\\
6&0.386&0.393&0.374&0.400&0.371&0.389&0.407&0.401&0.371&0.386\\
7&0.359&0.361&0.344&0.370&0.339&0.359&0.356&0.363&0.343&0.359\\
8&0.337&0.325&0.325&0.335&0.314&0.332&0.330&0.334&0.311&0.330\\
9&0.318&0.305&0.310&0.305&0.296&0.331&0.314&0.316&0.299&0.301\\
10&0.282&0.270&0.270&0.297&0.283&0.301&0.290&0.284&0.300&0.306\\
11&0.212&0.289&0.314&0.287&0.263&0.392&0.206&0.229&0.294&0.297\\
12&0.687&0.314&0.356&0.142&0.188&0.337&0.337&0.306&0.626&0.281\\
\hline
\end{tabular}
\caption {The values of $\Lambda_k$ for the Fake Guadalupe simulations for ten erandom circuits for $n=12$.} 
\label{tab:Figure7}
\end{center}
\end{table}

\subsection{Fourier behavior of the neutral atoms logical circuits (Figures 11,12)}

\begin{table}[H]\begin{center}
\begin{adjustbox}{max width=\textwidth}
 \begin{tabular}{||c c c c c c c c c c c c c||} 
 \hline
T/K&1&2&3&4&5&6&7&8&9&10&11&12\\[0.5ex] 
\hline\hline 
1&0.852&0.729&0.607&0.585&0.534&0.531&0.474&0.485&0.464&0.461&0.437&0.513\\
2&0.889&0.671&0.487&0.44&0.363&0.351&0.300&0.283&0.279&0.272&0.277&0.309\\
3&0.858&0.611&0.42&0.366&0.288&0.276&0.238&0.216&0.224&0.208&0.213&0.234\\
4&0.824&0.558&0.372&0.318&0.240&0.232&0.199&0.188&0.189&0.177&0.203&0.217\\
5&0.766&0.513&0.341&0.293&0.214&0.209&0.175&0.175&0.173&0.168&0.183&0.21\\
6&0.729&0.477&0.319&0.268&0.193&0.191&0.157&0.157&0.151&0.152&0.167&0.189\\
7&0.693&0.444&0.293&0.247&0.178&0.174&0.145&0.145&0.138&0.136&0.154&0.18\\
8&0.669&0.42&0.277&0.229&0.164&0.16&0.134&0.132&0.124&0.125&0.141&0.162\\
9&0.641&0.399&0.265&0.214&0.153&0.151&0.126&0.124&0.116&0.117&0.141&0.151\\
10&0.620&0.379&0.253&0.201&0.144&0.142&0.119&0.115&0.109&0.108&0.127&0.144\\
11&0.597&0.361&0.24&0.188&0.135&0.134&0.114&0.106&0.101&0.1&0.12&0.136\\
12&0.586&0.348&0.231&0.179&0.129&0.128&0.107&0.102&0.097&0.094&0.109&0.135\\
13&0.564&0.339&0.223&0.174&0.124&0.123&0.103&0.098&0.097&0.091&0.109&0.131\\
14&0.538&0.332&0.213&0.169&0.119&0.119&0.098&0.095&0.094&0.089&0.101&0.126\\
15&0.524&0.322&0.205&0.165&0.115&0.115&0.094&0.091&0.089&0.086&0.099&0.123\\
16&0.502&0.315&0.199&0.16&0.111&0.111&0.091&0.088&0.086&0.084&0.097&0.116\\
\hline
\multicolumn{11}{c}{} \\ \hline
T/K&1&2&3&4&5&6&7&8&9&10&11&12\\[0.5ex] 
\hline\hline 
1&1.009&0.834&0.801&0.731&0.680&0.673&0.621&0.613&0.599&0.585&0.518&0.523\\
2&0.952&0.736&0.617&0.545&0.465&0.463&0.400&0.393&0.380&0.378&0.350&0.325\\
3&0.917&0.679&0.527&0.454&0.369&0.367&0.308&0.295&0.285&0.278&0.255&0.247\\
4&0.882&0.620&0.464&0.397&0.312&0.311&0.256&0.251&0.242&0.235&0.225&0.217\\
5&0.828&0.571&0.426&0.364&0.280&0.279&0.234&0.230&0.220&0.217&0.202&0.209\\
6&0.780&0.529&0.395&0.335&0.254&0.255&0.210&0.206&0.202&0.191&0.182&0.188\\
7&0.759&0.499&0.368&0.309&0.231&0.230&0.189&0.191&0.183&0.175&0.171&0.187\\
8&0.738&0.470&0.347&0.285&0.211&0.213&0.170&0.173&0.163&0.159&0.152&0.180\\
9&0.705&0.447&0.329&0.270&0.200&0.200&0.159&0.163&0.153&0.150&0.138&0.167\\
10&0.685&0.426&0.311&0.255&0.188&0.187&0.151&0.155&0.146&0.146&0.137&0.159\\
11&0.667&0.407&0.294&0.241&0.178&0.178&0.143&0.147&0.137&0.134&0.129&0.146\\
12&0.649&0.390&0.282&0.231&0.170&0.169&0.136&0.141&0.131&0.126&0.121&0.136\\
13&0.622&0.381&0.272&0.224&0.164&0.163&0.132&0.136&0.127&0.118&0.116&0.131\\
14&0.600&0.373&0.261&0.218&0.158&0.158&0.128&0.130&0.122&0.115&0.111&0.127\\
15&0.582&0.362&0.251&0.212&0.153&0.153&0.123&0.126&0.118&0.114&0.110&0.124\\
16&0.561&0.352&0.243&0.205&0.148&0.149&0.118&0.122&0.114&0.109&0.105&0.122\\
\hline
\end{tabular}
\end{adjustbox}
\caption {The values of $\Lambda_k$ for the 16 samples based on different postselection for the first logical IQP circuit with 12 qubits (Figures 11,12).} 
\label{tab:Figure11}
\end{center}
\end{table}

\newpage

{\small

\noindent
Gil Kalai,  Hebrew University of Jerusalem, Einstein Institute of Mathematics, and\\  Reichman University, Efi Arazi School of Computer Science. \\ {\tt gil.kalai@gmail.com}.

\medskip

\noindent
Yosef Rinott,  Hebrew University of Jerusalem, Federmann Center for the Study of Rationality and Department of Statistics.\\ {\tt yosef.rinott@mail.huji.ac.il}.

\medskip

\noindent
Tomer Shoham,  Hebrew University of Jerusalem, Federmann Center for the Study of Rationality and Department of Computer Science.\\ {\tt tomer.shohamm@gmail.com}.

}

\end{document}